\documentclass[final,3p,times,onecolumn,number,review,reviewsort&compress]{elsarticle}
\usepackage{amssymb}
\usepackage{upgreek}
\usepackage{lineno}
\usepackage{hyperref}
\usepackage{makecell}
\usepackage{multirow}
\usepackage{caption}
\usepackage{subcaption}
\usepackage{graphicx}

\journal{Nuclear Instruments and Methods: A}

\begin{document}

\begin{frontmatter}

\title{Evaluation of the response of plastic scintillator bars and measurement of neutron capture time in non-reactor environment for the ISMRAN experiment}

\author[a,b]{R.~Dey\corref{cor1}}
\ead{ronidey@barc.gov.in}
\cortext[cor1]{Corresponding author}
\author[a]{P.~K.~Netrakanti}
\author[a]{D.~K.~Mishra}
\author[a]{S.~P.~Behera}
\author[a]{R.~Sehgal}
\author[a,b]{V.~Jha}
\author[b,c]{and L.~M.~Pant}

\address[a]{Nuclear Physics Division, Bhabha Atomic Research Centre, Trombay, Mumbai - 400085}
\address[b]{Homi Bhabha National Institute, Anushakti Nagar, Mumbai - 400094}
\address[c]{Technical Physics Division, Bhabha Atomic Research Centre, Trombay, Mumbai - 400085}

\begin{abstract}

We present a detailed study on detector response to different radioactive sources and the measurements of non-reactor environmental backgrounds with the Indian Scintillator Matrix for Reactor Anti-Neutrinos (ISMRAN) detector setup consisting of 9$\times$10 Plastic Scintillator Bars (PSBs) array at BARC, Mumbai. These measurements are useful in the context of the ISMRAN detector setup, which will be used to detect the reactor anti-neutrinos (${\overline{\ensuremath{\nu}}}_{e}$) and measure its energy spectra, through the inverse beta decay (IBD) process. A GEANT4 based Monte Carlo (MC) simulation is used to understand the optical transmission, energy resolution and energy non-linearity of the ISMRAN detector. A detailed analysis procedure has been developed to understand the natural radioactive, cosmogenic and cosmic muon-induced backgrounds with the ISMRAN detector setup in a non-reactor environment, based on their energy deposition, number of bars hit as well as topological event selection criteria in position and time for triggered events. Data and MC simulated distributions of reconstructed sum energy and number of bars hit has been compared for the radioactive $\gamma$ + positron source such as $\mathrm{{}^{22}Na}$ placed at the center of the ISMRAN array. Fast neutron energy response and capture time distribution in ISMRAN array has been studied using a novel technique involving Time of Flight (TOF) of the measured fast neutrons. The observed characteristic neutron capture time ( $\tau$ ) of 68.29 $\pm$ 9.48 $\mu$s is in good agreement with $\sim$65 $\mu$s obtained from MC simulation. These experimentally measured results will be useful for discriminating the correlated and uncorrelated background events from the true IBD events in reactor ON and OFF conditions inside the reactor hall.

\end{abstract}

\begin{keyword}


  Anti-neutrinos \sep Plastic scintillator \sep Reactor monitoring \sep Energy resolution \sep Energy non-linearity \sep Time of Flight.
\end{keyword}

\end{frontmatter}


\section{Introduction}
Reactor anti-neutrinos (${\overline{\ensuremath{\nu}}}_{e}$) have provided the key aspects for understanding of neutrinos, including the first observation of ${\overline{\ensuremath{\nu}}}_{e}$~\cite{Cowan} and the precise measurement of the neutrino mixing parameters $\mathrm{\theta_{13}}$, $\mathrm{\Delta m^2_{21}}$ and $|\mathrm{\Delta m^2_{31}}|$~\cite{PDG} in the three flavour framework. Experimental collaborations such as Daya Bay~\cite{DayaBay}, RENO~\cite{RENO} and Double Chooz~\cite{DChooz} have achieved detailed and precise measurements of ${\overline{\ensuremath{\nu}}}_{e}$ energy spectrum and $\theta_{13}$~\cite{DBay,Reno}. However, from the comparisons between the theoretical models and the experimental results, a $\sim$ 6 $\%$ deficit is observed in the global reactor ${\overline{\ensuremath{\nu}}}_{e}$ flux, which is known as ``reactor anti-neutrino anomaly" (RAA)~\cite{Mueller,Huber,Mention,Vogel}. A disagreement with the expected shape of reactor ${\overline{\ensuremath{\nu}}}_{e}$ induced prompt positron energy spectrum between 5$-$7 MeV  with a local significance of up to $\sim$4$\sigma$, has also been observed, which is known as ``shape distortion" or ``5 MeV bump"~\cite{DBay_bump,DChooz_bump}. The observed flux deficit RAA has led to the hypothesis of oscillations involving a sterile neutrino state with $\mathrm{\sim 1 eV^2}$ mass splitting or reasons of it may also lie in the complex nuclear physics of reactors~\cite{Sonzogni}. However, many ongoing effort for the search of eV-scale sterile neutrinos at very short baseline experiments ($\mathrm{\sim 10 m}$) have reported significant exclusions in the parameter space of mixing angle ($ \mathrm{sin^2(2\theta_{14})}$) and mass splitting ($\mathrm{\Delta m^2_{41}}$). Some of these experiments are also sensitive to monitor the reactor power in a non-intrusive way and estimate the fuel compositions as a function of burn up~\cite{SONGS,NUCIFER}. 

To explore these aspects, an array of plastic scintillator bars (PSBs) known as Indian Scintillator Matrix for Reactor Anti-Neutrinos (ISMRAN), has been installed and commissioned at Dhruva research reactor facility~\cite{DHRUVA} in Bhabha Atomic Research Centre (BARC), which uses natural uranium as fuel, at $\sim$13 m from the reactor core . It is designed to measure the yield and energy spectrum of ${\overline{\ensuremath{\nu}}}_{e}$, via the inverse beta decay (IBD) process. The measured spectra can be used for the searching of the sterile neutrino with a mass on the order of $\sim$1 eV /$\mathrm{c^{2}}$~\cite{NEOS,RENOST}. With the ISMRAN detector setup the active sterile neutrino oscillation can be observed with a 95$\%$ confidence level provided that $\mathrm{sin^{2}2\theta_{14}}$ $\geq$ 0.09 at $\mathrm{\Delta m^{2}_{41}}$ = 1 for an exposure of 1 ton-year~\cite{Shiba}. At the same time, the monitoring of the reactor thermal power and fuel evolution can be demonstrated using the measured ${\overline{\ensuremath{\nu}}}_{e}$ yields as a function of time~\cite{PROSPECT,DayaBayFuel,IAEA,Oguri}. The excess of ${\overline{\ensuremath{\nu}}}_{e}$ events in data compared to the predictions particularly between 5$-$7 MeV in the measured prompt positron energy spectrum will also be addressed using ISMRAN detector setup~\cite{DB5MeV,RENO5MeV,HUBER5MeV}. A first measurement of ${\overline{\ensuremath{\nu}}}_{e}$ candidate events with a prototype detector, mini-ISMRAN consisting of 1/16th of the original ISMRAN, was installed and successfully recorded data inside the Dhruva reactor hall in the year 2018. A total of 218 $\pm$ 50 (stat) $\pm$ 37 (sys) ${\overline{\ensuremath{\nu}}}_{e}$ candidate events were obtained for 128 days of reactor ON data~\cite{miniISMRAN}. The physics data collection has already started with the full scale ISMRAN detector setup at the end of the year 2021.

The ISMRAN detector setup, consists of an array of plastic scintillator bars (PSBs) arranged in 9$\times$10 (9 horizontals and 10 verticals) matrix and enclosed by a shielding made of 10 cm thick lead (Pb) and 10 cm thick borated polyethylene (BP) with 30$\%$ boron concentration, to minimize the influence of reactor backgrounds resulting from the photons and neutrons inside the reactor hall. From MC simulations it is estimated that a moderate shielding of 10 cm of Pb and 10 cm of BP has a rejection of up to $\sim88\%$ for the fast neutron events, within the energy range of 2 MeV to 10 MeV, inside the shielding structure ~\cite{ISMRAN}. This choice of shielding is optimized by keeping the tonnage limitation of the massive shielding structures on the floor of the reactor hall. The complete setup is mounted on a movable base structure which will allow us to make the measurements at different distances from the reactor core. Similar setup using an array of PSBs has been used by the PANDA experiment at a stand off distance of $\sim$25 m from the reactor core and have reported promising results~\cite{Oguri}. The DANSS collaboration has also demonstrated the performance of solid plastic strip scintillators for the reactor ${\overline{\ensuremath{\nu}}}_{e}$ detection~\cite{DANSS}.

Apart from reactor background, natural radioactive, cosmogenic and cosmic muon-induced backgrounds have also been encountered inside the reactor hall. Therefore a good understanding of cosmogenic background as well as natural radioactive background that constitutes the $\gamma$-rays mainly from $\mathrm{{}^{40}K}$ and $\mathrm{{}^{208}Tl}$ in plastic scintillator array is an essential pre-requisite to estimate their uncertainties in the measurements. This has been performed through a detailed measurement of non-reactor background with plastic scintillator array at the Detector Integration Laboratory (DIL) in BARC, without using any shielding structure in a non-reactor environment, which is situated about 1.5 km away from the Dhruva reactor hall.

In this paper, we present the results obtained from the comparison between data and GEANT4 based MC simulations which incorporate the optical transmission of the scintillation photons, energy resolution and energy non-linearity for PSB. We also discuss the response of the detector to natural radioactive, cosmogenic and cosmic muon-induced backgrounds, based on their sum energy distribution and number of bars hit for the triggered events inside the ISMRAN array. Reconstructed sum energy spectrum and number of bars hit for $\gamma$-rays + positron coming from radioactive source such as $\mathrm{{}^{22}Na}$ has been compared with MC simulations by placing it at the center of the ISMRAN array. Comparison between neutron capture on hydrogen (n-H) events and accidental background events have been studied in detail based on their sum energy and number of bars hit variable using a Am/Be neutron source, placed on the top of the ISMRAN array. The reconstructed $\sim$2.2 MeV $\gamma$-ray from n-H capture events from the cosmogenic neutron background or cosmic muon spallated secondary neutron will be useful for monitoring the energy scale stability of ISMRAN detectors inside the reactor hall. The sum energy, number of bars hit, individual PSB energy and energy deposition ratio variables are reconstructed for the cascade $\gamma$-rays from the neutron capture on Gadolinium (n-Gd) capture process using Am/Be neutron source and compared with the MC simulations in GEANT4. We also study the energy response of fast neutrons in PSBs, generated from Am/Be source using Time of Flight (TOF) technique where we can identify the fast neutron induced events from $\gamma$-ray events. These tagged fast neutrons from TOF are then used to measure the capture time distribution using a novel technique for the n-Gd capture events in the ISMRAN array and the results are compared with the MC simulations.

\section{Experimental setup and data acquisition system}
\begin{figure}[h]
\begin{center}
\includegraphics[width=14cm,height=7cm]{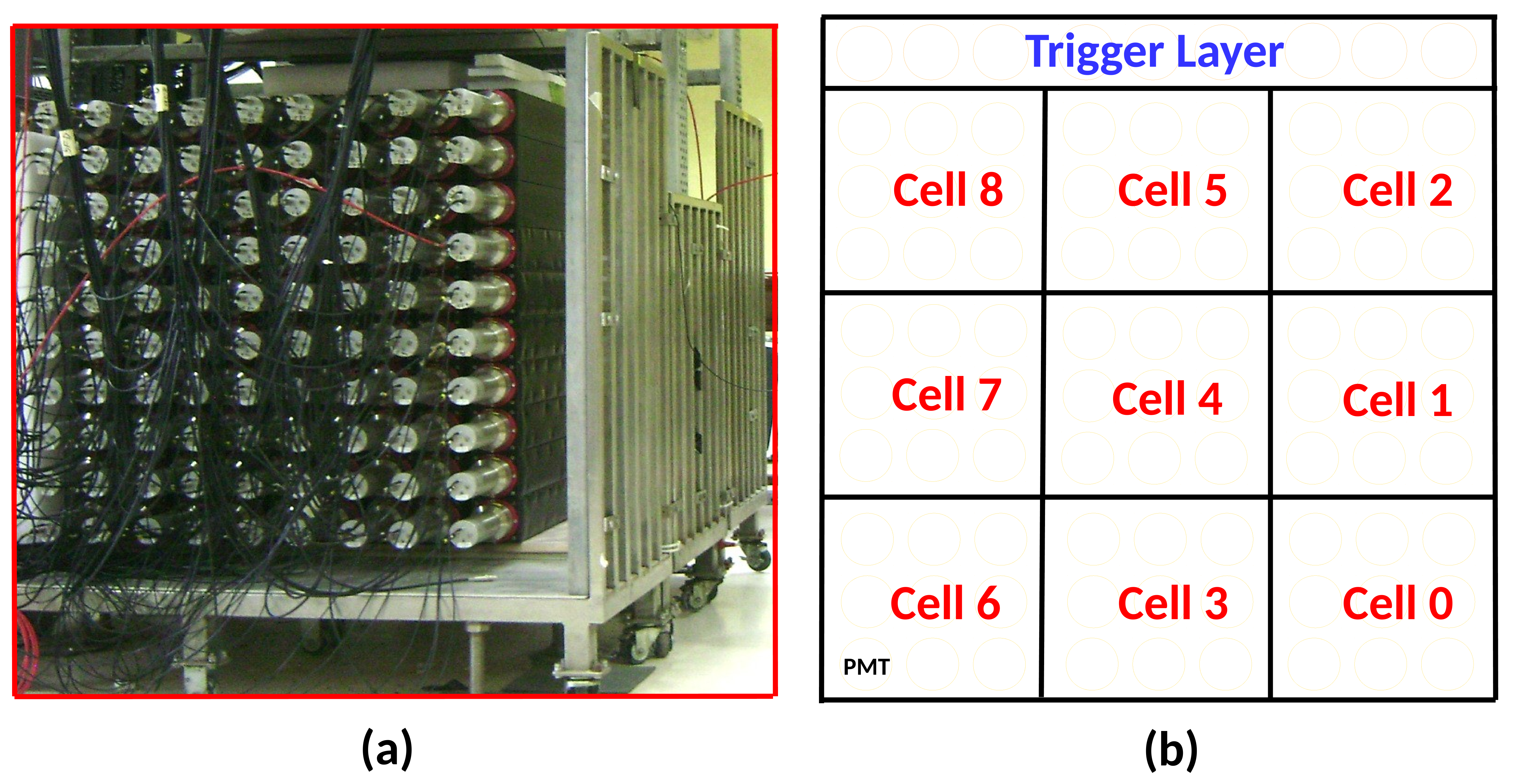}
\caption{Panel (a) shows the full scale ISMRAN detector setup consisting of 90 PSBs in DIL. Panel (b) shows the fiducial cell arrangement, each cell consisting of 9 PSBs, in ISMRAN array.}
\label{fig1}
\end{center}
\end{figure}

The detector setup at DIL, consists of 90 PSBs, arranged in the form of a matrix in an array of 9$\times$10. Each PSB is wrapped with Gadolinium Oxide ($\mathrm{Gd_{2}O_{3}}$) coated on aluminized mylar foils. The areal density of the $\mathrm{Gd_{2}O_{3}}$ on these foils is 4.8 $\mathrm{mg/cm^2}$ . Each PSB has a composition similar to EJ200~\cite{eljen} and is 100 cm long with a cross-section of 10 $\times$ 10 $\mathrm{cm^2}$. Three inch diameter, PMTs are coupled at the both ends of each PSB for signal readout of the triggered events. The complete detector setup approximately weighs $\sim$1 ton and is shown in Fig.~\ref{fig1} (a). 

 The data acquisition system (DAQ), a CAEN VME based 16 channels waveform digitizers (V1730) of 500 $\mathrm{MS/s}$, has been used for pulse processing and event triggering from each PSB independently. The pulse discrimination, trigger generation, threshold selection, charge integration in units of ADC, timestamp for each triggered event using constant fraction discrimination (CFD) and coincidence of the PMT signals for each PSB are done using on-board FPGAs of the digitizers. The anode signals from the PMTs at both ends of a PSB are required to have a timing coincidence of 36 ns to be recorded as a triggered event. The timestamped data from each PSB is then further analyzed offline using energy deposition, timestamps and position information to build an event. To collect data from all 90 PSBs independently, with practically zero dead time, 12 CAEN VME based waveform digitizers (V1730) have been used in a time synchronized mode. 

\section{Antineutrino detection principle through IBD}
The reactor based ${\overline{\ensuremath{\nu}}}_{e}$ events in ISMRAN array are detected using the inverse beta decay (IBD) process, which has an interaction cross-section of approximately $\mathrm{6 \times 10^{-43} cm^{2}}$~\cite{Mention,Vogel}. The interaction of ${\overline{\ensuremath{\nu}}}_{e}$ with protons present in PSB generates a positron and a neutron as daughter particles via IBD process, as shown in Eq.~\ref{eq:ibd}.
\begin{equation}\label{eq:ibd}
\mathrm{ \overline \nuup_{e} + p \rightarrow e^{+} + n}.
\end{equation}
The minimum ${\overline{\ensuremath{\nu}}}_{e}$ energy required for the above reaction to occur is $\sim$1.8 MeV. The positron being lighter in mass as compared to neutron, carries most of the ${\overline{\ensuremath{\nu}}}_{e}$'s energy, leaving a few keV's of energy for the neutron. The positron looses energy, through ionization in the PSB followed by annihilation with a surrounding electron, resulting in emission of two $\gamma$-rays of energy 0.511 MeV each. The signal produced by the energy loss of positron and the Compton scattered $\gamma$-rays from the annihilation comprise the prompt event signature.
On the other hand, the neutron which has only a few keV’s of energy, loses its energy through multiple scattering and eventually thermalize in the scintillator volume and gets captured either in Gadolinium (Gd) foils or in Hydrogen (H) present in the scintillator volume. 

\begin{figure}[h]
\begin{center}
  \includegraphics[width=8cm,height=7.4cm]{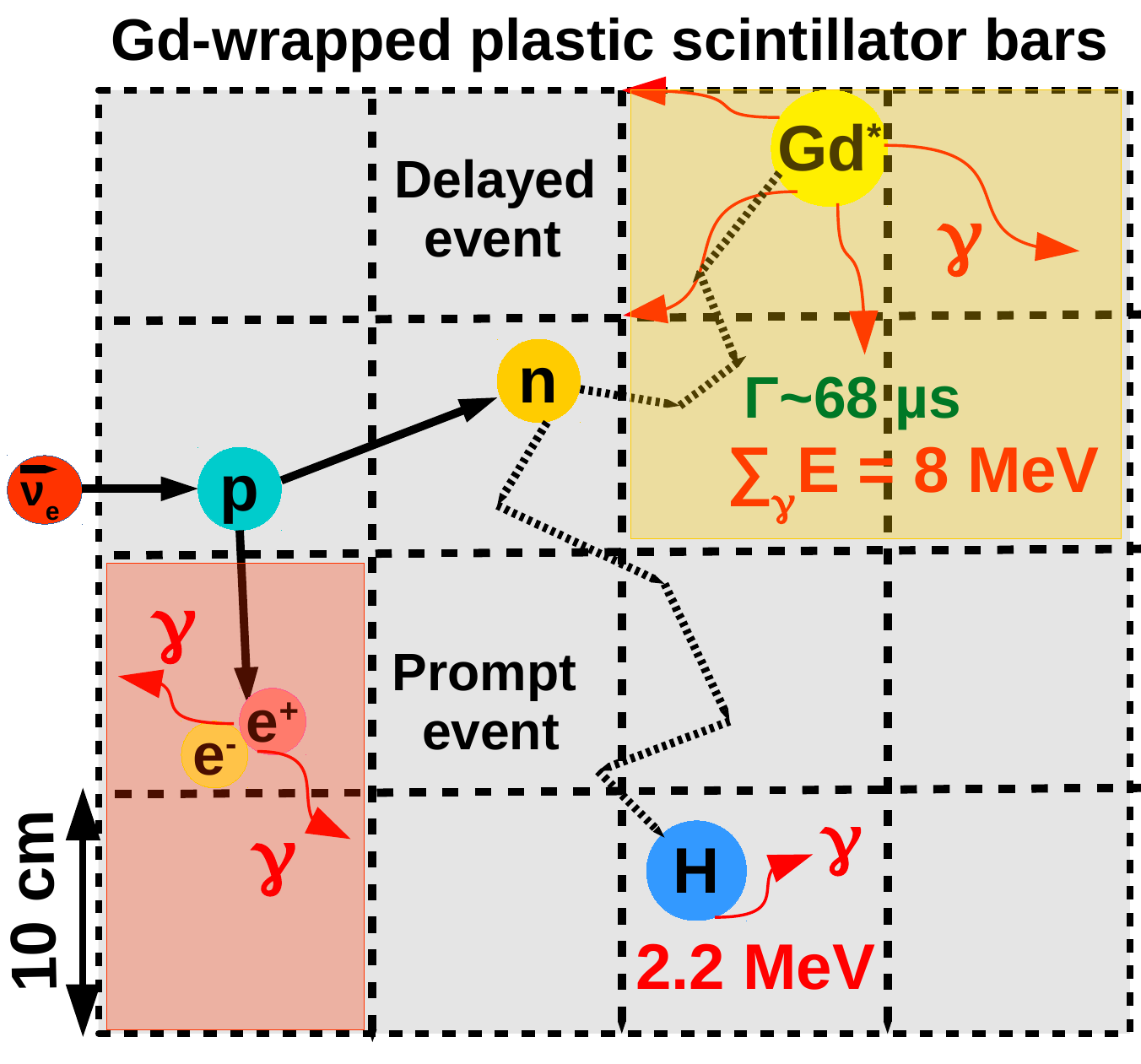}
\caption{ Schematic representation of IBD event generating prompt and delayed event signatures in mini-ISMRAN array.}
\label{fig2}
\end{center}
\end{figure}

\begin{equation}\label{eq:Gd1}
  \hspace{-0.2in}
\mathrm{ n + {}^{155}Gd \rightarrow {}^{156}Gd^{*} \rightarrow \gamma 's,   \quad \sum E_{\gamma} = 8.5~MeV}, \quad \sigmaup_{\mathrm{n-capture}} = \mathrm{61000~b},
\end{equation}
\begin{equation}\label{eq:Gd2}
  \hspace{-0.2in}
\mathrm{ n + {}^{157}Gd \rightarrow {}^{158}Gd^{*} \rightarrow \gamma 's,  \quad \sum E_{\gamma} = 7.9~MeV}, \quad \sigmaup_{\mathrm{n-capture}} = \mathrm{254000~b}.
\end{equation}
Due to the larger thermal neutron capture cross-section for Gd nuclei, as shown in Eqs.~\ref{eq:Gd1} and ~\ref{eq:Gd2}, from simulations it has been observed that $\sim$75$\%$ of the events are captured on Gd nuclei and rest $\sim$25$\%$ events are captured on H nuclei in ISMRAN detector geometry~\cite{ISMRAN}. The neutron capture on H nuclei emitting a mono-energetic $\gamma$-ray of energy 2.2 MeV is difficult to tag as a delayed partner due to the presence of reactor related $\gamma$-ray background in that energy region.
However, the de-excitation of the Gd nucleus leads to emission of cascade $\gamma$-rays of sum energy $\sim$8 MeV, forms much robust delayed event signature.
The sum energy depositions for the prompt positron and delayed neutron capture events along with the selection of number of bars hit uniquely identifies the  ${\overline{\ensuremath{\nu}}}_{e}$ candidate events in ISMRAN array, as shown in Fig.~\ref{fig2}. Also, the characteristic time difference ($\mathrm{\tau}$) between the prompt and the delayed event signatures, is of the order of $\sim$68 $\mathrm{\mu}$s for the ISMRAN detector geometry obtained from MC simulations, which can discriminate the ${\overline{\ensuremath{\nu}}}_{e}$ from other correlated and uncorrelated background events~\cite{ISMRAN}.

\section{ISMRAN detector response}
\begin{figure}[h]
\hspace{-2.9em}  
\includegraphics[width=18cm,height=5.0cm]{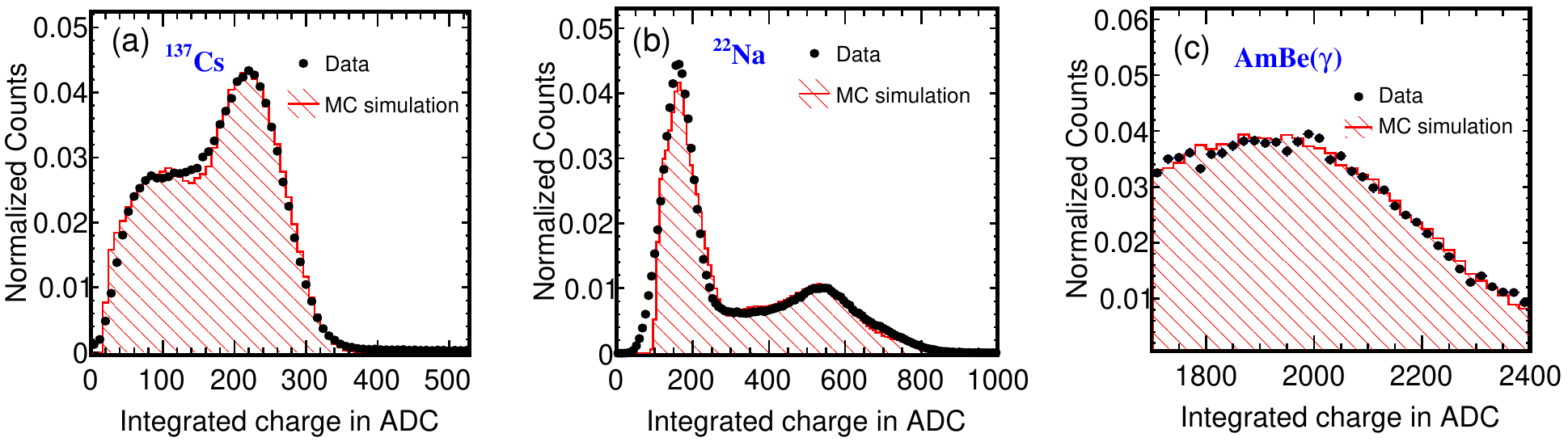}
\caption{Panels (a), (b) and (c) show the comparisons of integrated charge spectra, in a single PSB, between data and MC simulated events for different radioactive sources such as $\mathrm{{}^{137}Cs}$, $\mathrm{{}^{22}Na}$ and $\gamma$-ray from Am/Be, respectively. }
\label{fig3}
\end{figure}

A good understanding of the response of the PSB is an essential pre-requisite for the analysis presented in this paper. In MC simulations, based on GEANT4~\cite{GEANT4}, the standard electromagnetic and radioactive decays physics processes are used for the response of $\gamma$-rays, electrons and positrons from radioactive sources. The high precision QGSP BIC HP physics processes is used for the low energy neutron interactions in GEANT4 based MC simulation. By default GEANT4 package used photon evaporation model to simulate cascade $\gamma$-rays emission from the thermal neutron capture on Gd nucleus. However, in photon evaporation model the final sum energy of the captured event is conserved, but event-by-event individual cascade $\gamma$-rays distribution is not properly modeled. To get an accurate description of cascade $\gamma$-rays emission from the thermal neutron capture on Gd in our simulation studies, we have used the DICEBOX package which models the emission of cascade $\gamma$-rays from the region of high energy level density in an excited nucleus, based on the different models of energy level density and photon strength functions~\cite{DICEBOX,ANRIGD}. To obtain the uniform response of the measured signal in data among all the PSBs, a gain matching of PMTs is performed and the energy, timing and position characterization of each PSB is done independently. The middle of each PSB is considered as the center position at 0 cm and both extreme ends are taken as -50 cm and +50 cm, respectively. Using the timestamp information from both the PMTs coupled at the ends of each PSB, we can reconstruct the timing difference ($\mathrm{\Delta T_{LR}}$) and parameterize an approximate position (Zpos) by placing a $\mathrm{{}^{137}Cs}$ source at 9 different positions, from -40 cm to +40 cm with the step size of 10 cm, along the length of the PSB. Using this parametrization we can reconstruct the hit position distribution along the length of the PSB. The measured timing resolution of $\sim$4 ns leads to a parameterized position resolution of $\sim$20 cm in a single PSB~\cite{ISMRAN,RAMAN} and is almost uniform among the different bars. Due to the segmented geometry of the ISMRAN detector, we can exploit the timing and position information for the localization of an event in the ISMRAN array to differentiate an ${\overline{\ensuremath{\nu}}}_{e}$ event from a background event arising due to fast neutron~\cite{MLP,Roni}.
The complex absorption and re-emission processes of optical photons are known to be an important source of the non-linear and non-uniform response of the charge deposition  in the PSBs. Therefore, a good understanding of light propagation in PSB is desirable to model the detector response and reduce the systematic errors. In order to get the optical model for ISMRAN detectors, optical photons have been simulated, by placing different radioactive sources such as $\mathrm{{}^{137}Cs}$, $\mathrm{{}^{22}Na}$ and $\gamma$-rays from Am/Be sources at the center of the PSB, using optical physics process in GEANT4. The optical photons inside the scintillator volume are then propagated to the PMTs and the digitized output are then compared with the measured integrated charge spectra obtained from the data. Figure~\ref{fig3} (a), (b) and (c) show the comparisons between data and MC for integrated charge spectrum from all the three radioactive sources, namely $\mathrm{{}^{137}Cs}$, $\mathrm{{}^{22}Na}$ and $\gamma$-ray from Am/Be, respectively. A reasonable agreement between data and MC simulated events for the integrated charge spectrum in PSB from radioactive $\gamma$-ray sources are observed with an average $\chi^{2}$ of the order of $\sim$0.4 in each case. A small differences, between data and MC generated events, observed in $\mathrm{{}^{137}Cs}$ and $\mathrm{{}^{22}Na}$ spectra at lower integrated charge range are mainly due to the absence of natural background component and the  implementation of the individual PSB thresholds in the MC generated simulated events. 


To obtain the energy resolution in PSB, the energy deposition ($\mathrm{E_{bar}}$) has been simulated for different radioactive sources, such as $\mathrm{{}^{137}Cs}$, $\mathrm{{}^{22}Na}$ and $\gamma$-ray from Am/Be source, in a single PSB using radioactive decay physics process in GEANT4. 
The simulated $\mathrm{E_{bar}}$ is then smeared using a Gaussian approximation whose $\mathrm{\sigma}$ value are obtained from the energy dependent resolution function as shown in eq~\ref{eq:Resolution}.
\begin{equation}\label{eq:Resolution}
  \hspace{-0.2in}
\mathrm{ Resolution \left (\frac{\sigma_{E}}{E_{bar}} \right ) = \sqrt{A^{2} + \frac{B^{2}}{E_{bar}} + \frac{C^{2}}{E_{bar}^{2}}}},
\end{equation}

\begin{figure}[h]
\hspace{-2.9em}  
\includegraphics[width=18cm,height=5.4cm]{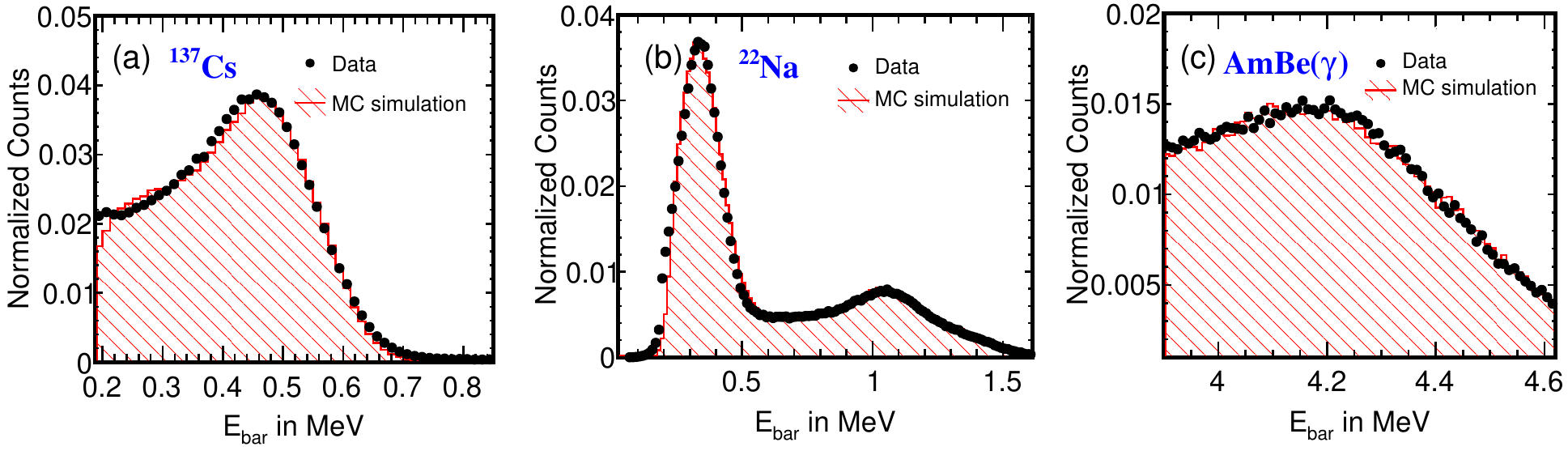}
\caption{ Panels (a), (b) and (c) show the comparisons of the calibrated $\mathrm{E_{bar}}$ distribution, in a single PSB,  between data and MC simulation from $\mathrm{{}^{137}Cs}$, $\mathrm{{}^{22}Na}$ and $\gamma$-ray Am/Be radioactive sources, respectively.}
\label{fig4}
\end{figure}

where the first term (A) is dependent on the light collection inefficiency variations, the second term (B) represents the energy dependent photo-statistics, and the third term (C) is related to PMT and electronics noise~\cite{stereo,daya_bay}. Figure~\ref{fig4} (a), (b) and (c) show the comparisons between measured $\mathrm{E_{bar}}$ spectra from data and MC simulated events for $\mathrm{{}^{137}Cs}$, $\mathrm{{}^{22}Na}$ and $\gamma$-rays from Am/Be source, respectively. The comparisons of $\mathrm{E_{bar}}$ between the data and MC events of a single PSB for different radioactive sources are reasonably in good agreement and the $\chi^{2}$ are tabulated in table~\ref{table1}. The disagreement between data and simulated results at lower energy range are mainly due to the absence of natural background component in the simulation results. 

\begin{table}[h]
\begin{small}
  \begin{center}
  \caption{Comparisons of Compton edges (CE) of $\mathrm{E_{bar}}$ between the data and MC results for different radioactive sources in a single PSB. Peak energies is given in MeV. Statistical errors of the fit results are negligible.}
  \label{table1}
\begin{tabular}{|c|c|c|c|}
\hline
\makecell{Radioactive sources} & $\mathrm{E_{data}}$ (MeV) & $\mathrm{E_{mc}}$ (MeV) & $\mathrm{{\chi}^2}$  \\
\hline
\makecell{$\mathrm{{}^{137}Cs}$} ~[CE = 0.478~MeV] & 0.466 & 0.464 & 0.56\\
\hline
\makecell{$\mathrm{{}^{22}Na}$}~[CE = 0.341~MeV]  & 0.350 & 0.340 & 0.75 \\
\hline
\makecell{$\mathrm{{}^{22}Na}$}~[CE = 1.060~MeV]  & 1.058 & 1.052 & 0.1 \\
\hline
\makecell{$\mathrm{AmBe~[CE = 4.196~ MeV]}$} & 4.20 & 4.198 & 0.17 \\
\hline
\end{tabular}
\end{center}
\end{small}
\end{table}

\begin{figure}[h]
\hspace{1.8em}  
\includegraphics[width=13cm,height=5.6cm]{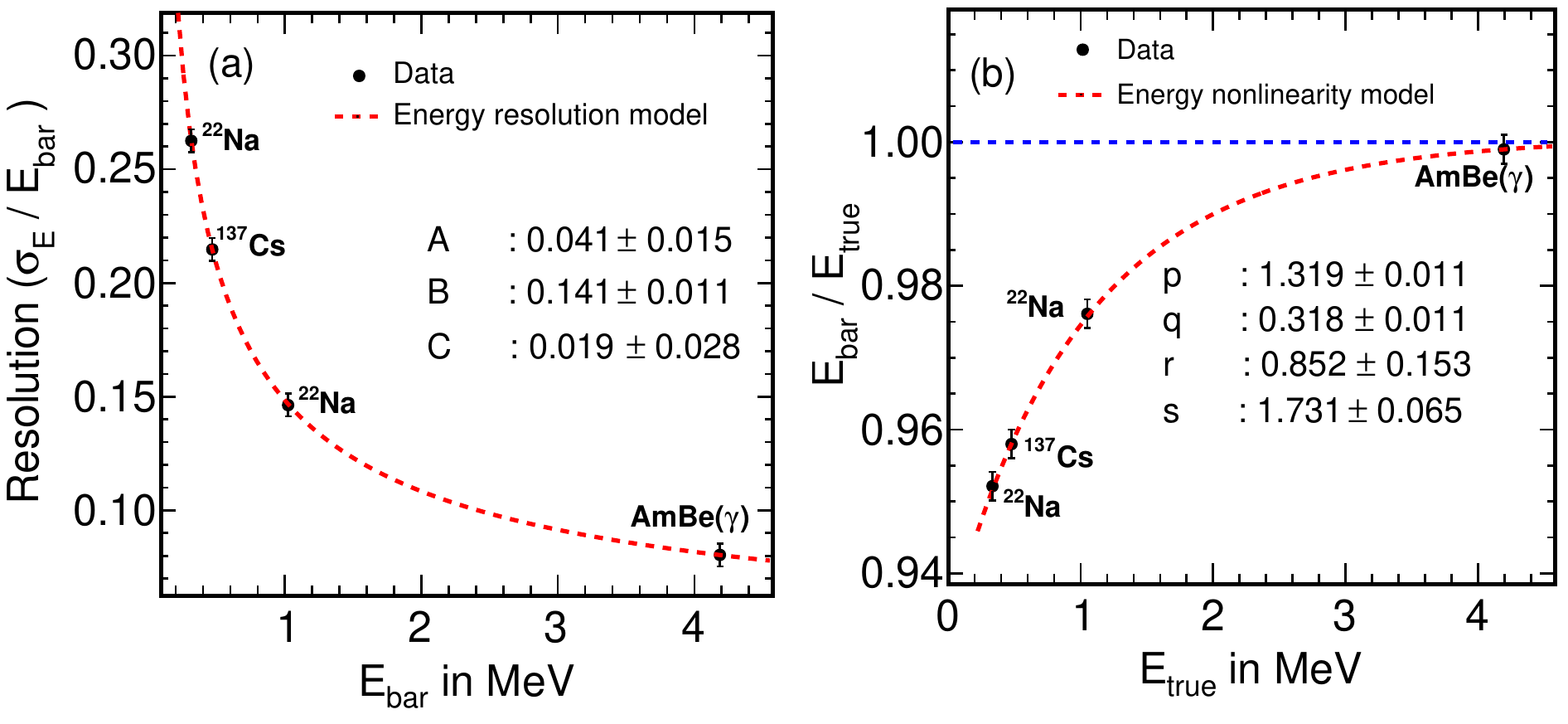}
\caption{ Panel (a) shows the energy resolution as a function of reconstructed $\gamma$-ray energy and panel (b) shows the non-linear response of scintillating energy in PSB obtained from the true Compton energies of $\gamma$-rays originating from different radioactive sources.}
\label{fig5}
\end{figure}

Figure~\ref{fig5} (a) displays the energy dependent resolution model with reconstructed $\gamma$-ray energy for $\mathrm{{}^{137}Cs}$, $\mathrm{{}^{22}Na}$ and $\gamma$-rays from Am/Be source in a single PSB. These results indicate that energy resolution of ISMRAN is well described by the standard resolution formula, which is shown in  Eqs.~\ref{eq:Resolution}, and that the photo-statistics term is the primary determinant of achieved energy resolution for ISMRAN, which is $\sim$14$\%$/$\sqrt{\mathrm{E_{bar}}}$.
Figure~\ref{fig5} (b) shows the nonlinear response of scintillating energy in a single PSB for different radioactive $\gamma$-rays sources, that is well described by a fitted parametrization and consistent with the MC predictions. The nonlinear response at lower energies is mainly due to the quenching effect in the scintillator volume~\cite{stereo,daya_bay}. The following empirical formula as shown in Eq.~\ref{eq:Nonlinearity} is used for the description of the energy non-linearity of PSB,

\begin{equation}\label{eq:Nonlinearity}
\mathrm {\frac{E_{bar}}{E_{true}} = p - \frac{q}{[1 - e^{-(rE_{true} + s)}]}}
\end{equation}

where $\mathrm{E_{true}}$ is true energy, in MeV, representing the Compton energy of $\gamma$-rays originating from different radioactive sources and $\mathrm{E_{bar}}$ corresponds to the reconstructed energy in a single PSB. The fit parameters p determines the saturation level, q corresponds to the magnitude of non-linearity, r and s are related to the shape of the non-linearity~\cite{daya_bay}.

\begin{figure}[h]
\hspace{-2.9em}  
\includegraphics[width=18cm,height=5.4cm]{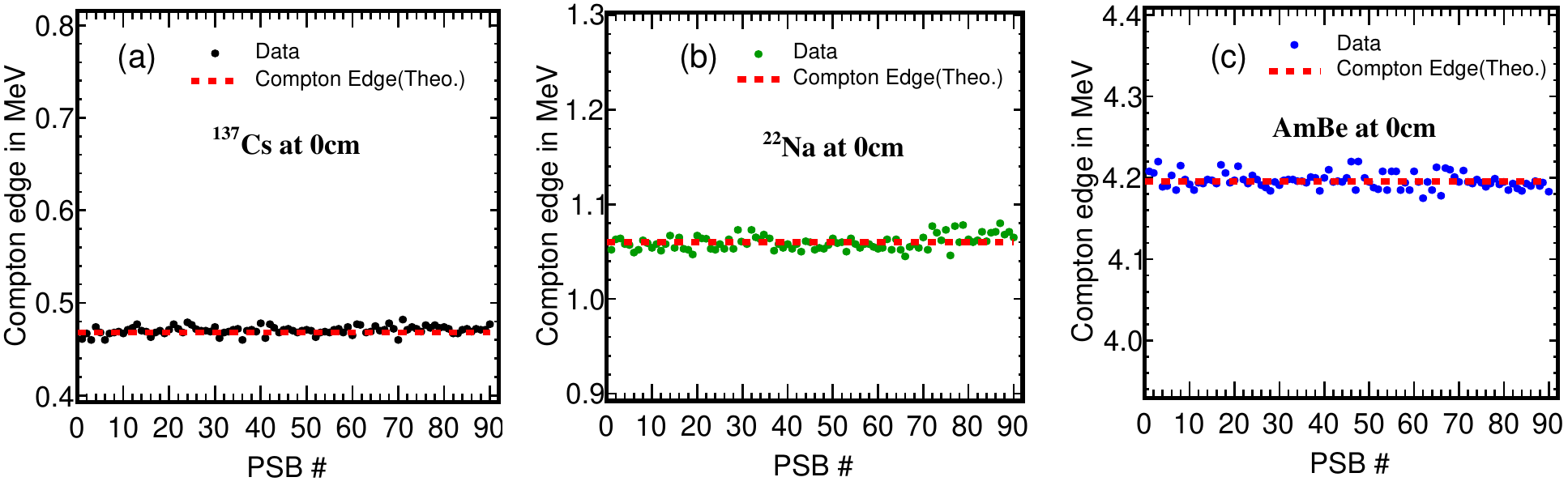}
\caption{ Panels (a), (b) and (c) show calibrated Compton edges among the 90 PSBs for different radioactive sources $\mathrm{{}^{137}Cs}$, $\mathrm{{}^{22}Na}$ and Am/Be, respectively.}
\label{fig6}
\end{figure}

The $\mathrm{E_{bar}}$ is obtained by calibrating each PSB using $\gamma$-rays and their respective Compton edges from radioactive sources such as $\mathrm{{}^{22}Na}$ and $\mathrm{{}^{137}Cs}$ placed at the center of the PSB. For higher energy calibration point we have used the 4.4 MeV $\gamma$-ray from the Am/Be source. Figure~\ref{fig6} (a),(b) and (c) show the Compton edges among all the 90 PSBs for $\mathrm{{}^{137}Cs}$, $\mathrm{{}^{22}Na}$ and Am/Be, respectively. Uniformity in the calibrated energy deposition is observed among all the 90 PSBs for different energy ranges with a variation of $\sim$3$\%$ among bars.

\section{Cosmogenic and natural radioactive background measurements with the ISMRAN setup}
One of the dominant source of background for the detection of ${\overline{\ensuremath{\nu}}}_{e}$ events in an above ground reactor ${\overline{\ensuremath{\nu}}}_{e}$ experiment is the presence of the interfering cosmic muon-induced and natural radioactive background. Before the reconstruction of the ${\overline{\ensuremath{\nu}}}_{e}$ events, all the events which satisfies the prompt and delayed event selection criteria are termed as prompt-like and delayed-like events, respectively. 
Two main background sources for the ${\overline{\ensuremath{\nu}}}_{e}$ candidate events arise from the uncorrelated and correlated pairs of prompt-like and delayed-like events. The uncorrelated background is due to the accidental coincidences from the random association of a prompt-like event due to natural radioactivity, with a delayed-like neutron capture event from background. The sources of prompt-like background events are from the $\gamma$-rays from the radioactivity in the PMT glasses, PSB, and  high energy $\gamma$-rays emanating from the neutron captures on the surrounding material, mostly iron, inside the reactor hall~\cite{ReactorBkg}. The ambient radioactivities generate $\gamma$-rays ranges in energy below 3 MeV. The delayed-like background events come from the neutrons produced by cosmic muons in the surrounding walls, shielding material, or inside the ISMRAN detector. The correlated ${\overline{\ensuremath{\nu}}}_{e}$ background events are due to the fast neutrons, $\beta$-n emitters from cosmogenic $\mathrm{{}^{9}Li}$/$\mathrm{{}^{8}He}$ isotopes and radioactive isotopes such as $\mathrm{{}^{12}B}$/$\mathrm{{}^{12}N}$~\cite{B12}. An energetic neutron entering the ISMRAN array can mimic the prompt-like event signature by proton recoil energy deposition and followed by thermalization and capture on Gd providing the delayed-like event signature. By exploiting the nature of the energy deposition profile in PSBs for the measured fast neutrons from D$-$D and D$-$T reactions and simulated prompt event from ${\overline{\ensuremath{\nu}}}_{e}$, a separation of these fast neutron background events are demonstrated in ref~\cite{Roni}.

\begin{figure}[h]
\begin{center}
\hspace{-2.9em}  
\includegraphics[width=15cm,height=6.0cm]{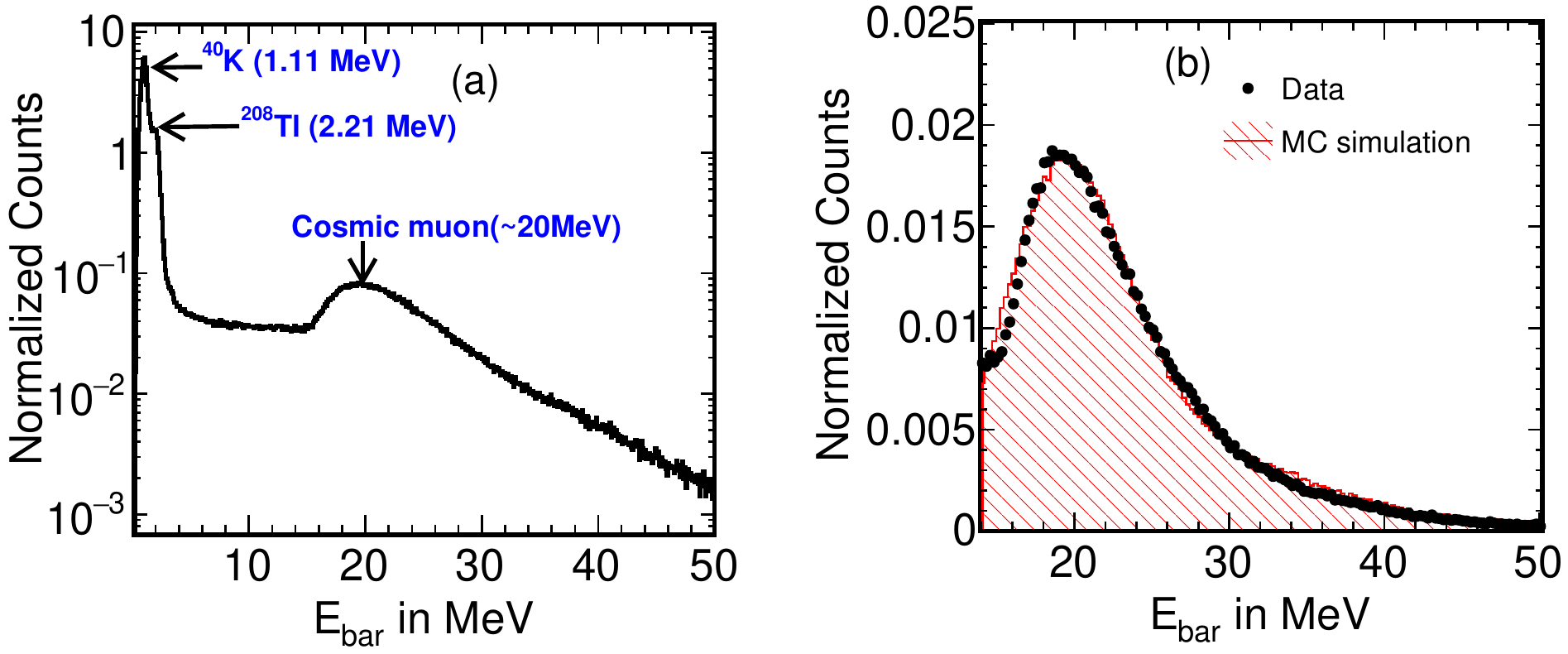}
\caption{ Panel (a) shows the calibrated $\mathrm{E_{bar}}$ distribution in a single PSB for the non-reactor background and panel (b) shows the comparison of the calibrated $\mathrm{E_{bar}}$ distribution between data and MC simulated events for the cosmic muon in a single PSB.}
\label{fig7}
\end{center}
\end{figure}

In order to understand and discriminate the non-reactor background, a detailed measurement of the non-reactor background with ISMRAN array has been accomplished at the DIL in BARC, without using any shielding structure. The characterization of different non-reactor backgrounds are done on the basis of cell arrangement, consisting of 9 PSBs in each cell, as shown in Fig.~\ref{fig1} (b). These cells configuration are implemented in the offline analysis only and not in the trigger while taking data. The 9 cells arrangement in the analysis is particularly chosen keeping the containment of the prompt and delayed event containment to the optimal level, as obtained from MC simulation studies. The top most layer consisting of 9 PSBs are used as a trigger layer for cosmic muon events as shown in Fig.~\ref{fig1} (b). The time interval ($\Delta T$) distribution of two consecutive muons has been evaluated in the triggered layer of ISMRAN array at sea level, which is demonstrated in detail in ref~\cite{cosmic}. Figure~\ref{fig7} (a) shows the $\mathrm{E_{bar}}$ in a single PSB due to the non-reactor backgrounds. A feature at $\sim$ 20 MeV is due to the minimum ionization energy deposition by cosmic muons in the 10 cm thick PSB. Natural radioactive background mainly dominated by the Compton edges of $\mathrm{{}^{40}K}$ ($\mathrm{E_{bar}}$ = 1.11 MeV) and $\mathrm{{}^{208}Tl}$ ($\mathrm{E_{bar}}$ = 2.21 MeV) are below 3 MeV. Depending on their energy deposition in PSB, non-reactor backgrounds has been separated into two regions, (A) natural radioactive background region $\mathrm{(0.2 < E_{bar} (MeV) < 10.0)}$ and (B) cosmic muon background region $\mathrm{(15.0 < E_{bar} (MeV) < 50.0)}$. Figure~\ref{fig7} (b) shows the comparison between the $\mathrm{E_{bar}}$ from data and MC simulated events for cosmic muon in a single PSB. There is a reasonable agreement between data and MC results for single bar $\mathrm{E_{bar}}$ for cosmic muon events and the energy region between 15.0 MeV to 50.0 MeV can be used for triggering of the cosmic muon events in ISMRAN array. 

\begin{figure}[h]
\begin{center}
\hspace{-2.9em}  
\includegraphics[width=14cm,height=6.2cm]{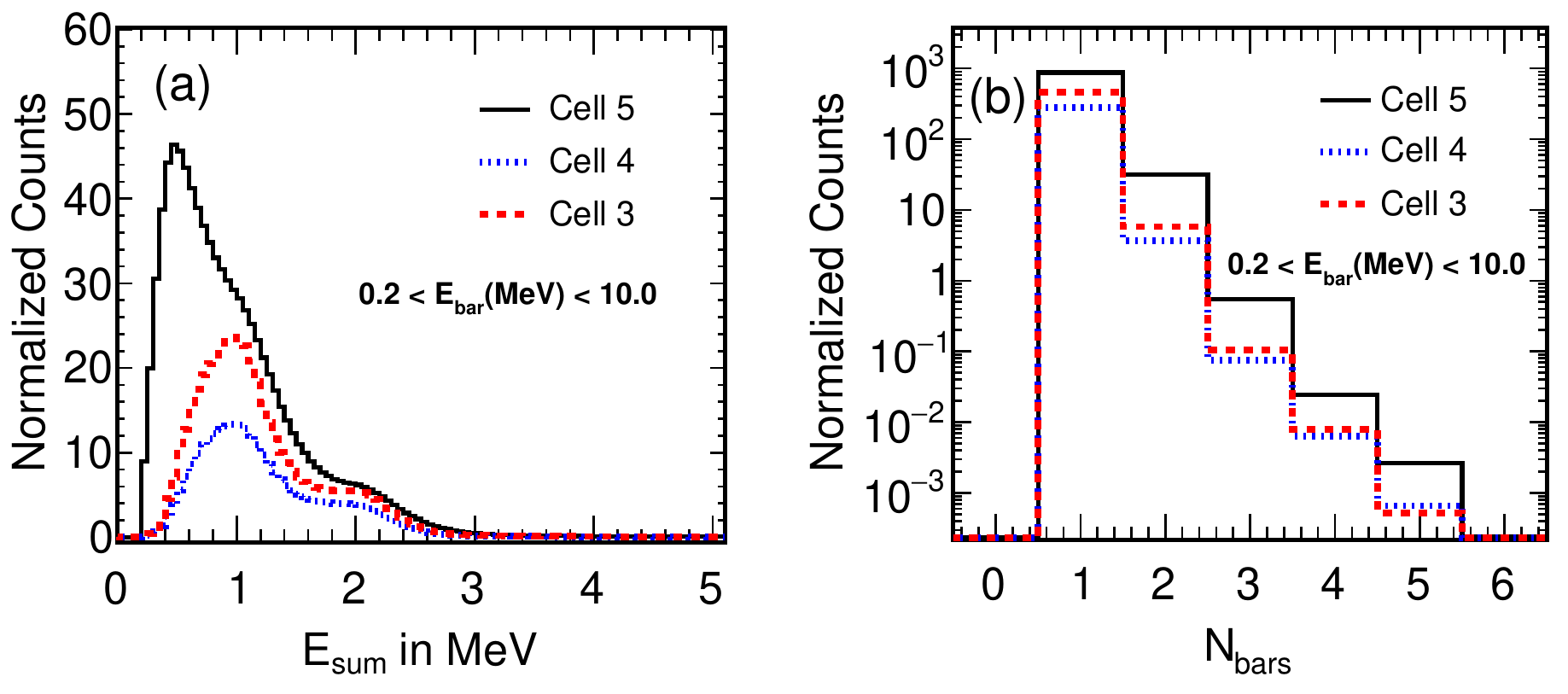}
\caption{ Panel (a) shows cell-wise reconstructed sum energy ($\mathrm{E_{sum}}$) and panel (b) shows the number of bars hit ($\mathrm{N_{bars}}$) distribution for natural radioactive background region.}
\label{fig8}
\end{center}
\end{figure}
\begin{figure}[h]
\begin{center}
\hspace{-2.9em}  
\includegraphics[width=14cm,height=6.0cm]{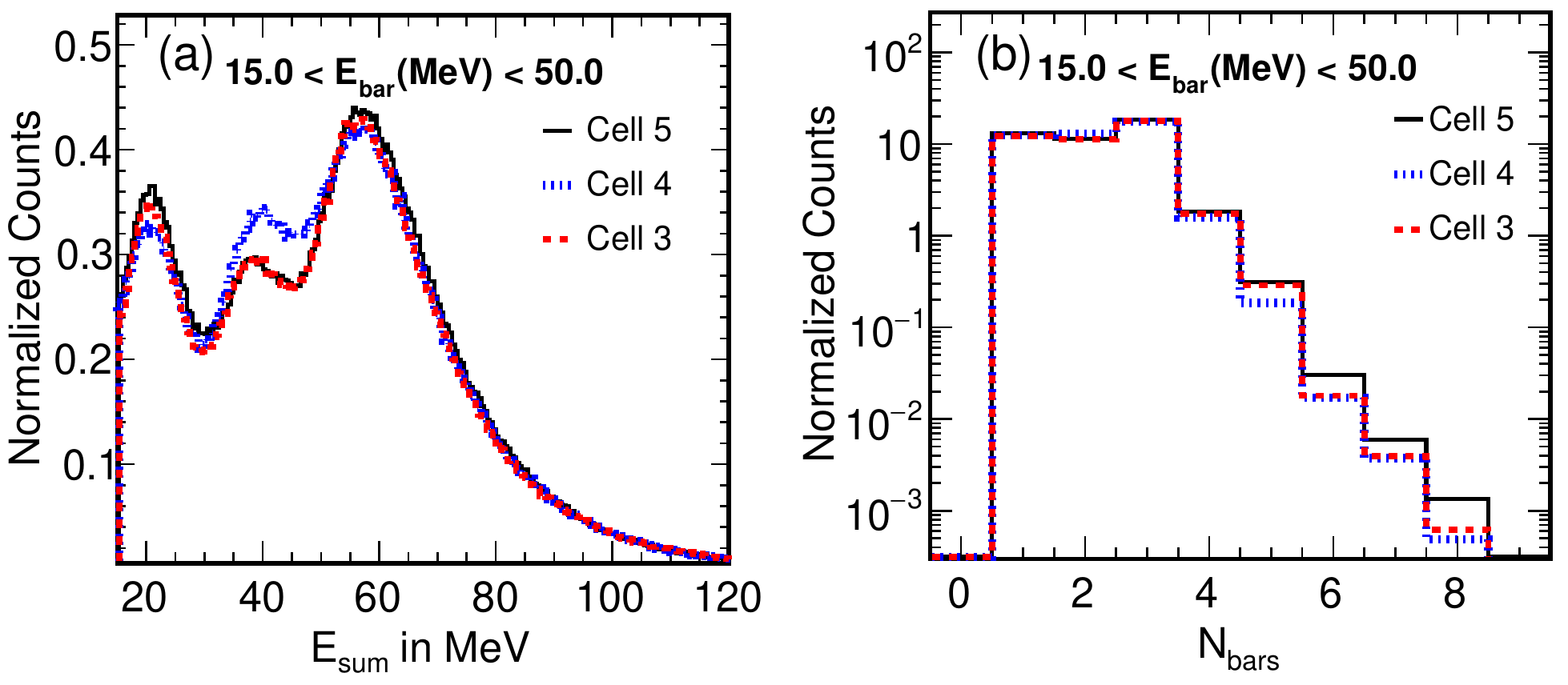}
\caption{Panel (a) and (b) show the cell-wise reconstructed sum energy ($\mathrm{E_{sum}}$) and number of bars hit ($\mathrm{N_{bars}}$) distribution for cosmic muon background, respectively.}
\label{fig9}
\end{center}
\end{figure}

The sum energy $\mathrm{(E_{sum})}$ and the number of bars hit $(\mathrm{N_{bars}})$ variables are constructed, in each cell, for natural radioactive and cosmic muon background regions for different cells of ISMRAN array. This scheme is used for the estimation of the position dependent background in PSBs for the non-reactor environmental background. For this study, as shown in Fig.~\ref{fig1} (b), we are showing only the comparison results from the cell 3, 4 and 5 which are at the center of the ISMRAN array. Similar results are observed in other cell combinations.
$\mathrm{E_{sum}}$ is calculated using the PSBs which have energy deposition above 0.2 MeV.  For calculation of $\mathrm{E_{sum}}$ for a cell, we take only those PSBs which are assigned in the cell from Fig.~\ref{fig1} (b) and do the arthematic sum of the energy deposition in each PSB corresponding to that cell.
Figure~\ref{fig8} (a) and (b) show the comparative study of the $\mathrm{E_{sum}}$ and the $\mathrm{N_{bars}}$ distributions for natural radioactive background region among different cells, respectively. From Fig.~\ref{fig8} (a), it can be seen that the dominant background up to 3 MeV is from the natural radioactivity and mainly from $\mathrm{{}^{40}K} ( E_{bar}$ = 1.11 MeV) and $\mathrm{{}^{208}Tl} ( E_{bar}$ = 2.21 MeV ). The yield of this background in cell 4, which is the center most cell, is less as compared to other two cells indicating the qualitative enhanced purity for ${\overline{\ensuremath{\nu}}}_{e}$ events can be obtained in this region of the ISMRAN array. This also shows the uniformity of the energy scale across the PSBs for the reconstruction of the correlated events in ISMRAN array.
Similar feature is also observed in the  $\mathrm{N_{bars}}$ distribution as shown in Fig.~\ref{fig8}(b).
Figure~\ref{fig9} (a) and (b) show the comparative study of $\mathrm{E_{sum}}$ and $\mathrm{N_{bars}}$ distributions for cosmic muon background region, respectively.
The $\mathrm{E_{sum}}$ distribution shows three distinct peaks, around 20 MeV, 40 MeV and 60 MeV,  which corresponds to the passage of muons in one, two and three bars in each cell, respectively. Enhanced event rate at 60 MeV compare to 20 MeV and 40 MeV indicates that most of the high energy cosmic muons passing through the cells vertically. 
Atmospheric muons with an average energy of $\sim$4 GeV passing through the scintillator bar covering a distance of $\sim$10 cm approximately looses 20 MeV of energy in a single PSB. These high energy muons act as minimum ionizing particle with nearly constant energy deposition in PSBs. Since the cell 4 is the innermost cell, covered all around by detectors, a passing muon has to deposit atleast this minimum energy in two PSBs. This is not the case for other cells, where there can be passing muons which can span any number of PSBs to come out of the cell. Hence there is an increase in the yield at 40 MeV for the center cell 4 as compared to the top cell 5 and the bottom cell 3. 
In this way the muon contamination in the ISMRAN array can be quantified in the reactor data and the muon induced background events can be circumvented for the analysis of ${\overline{\ensuremath{\nu}}}_{e}$ events. Figure~\ref{fig9} (b) shows the $\mathrm{N_{bars}}$ distribution for three cells and can be seen for cell 4 the number of bars fired at 2 is higher as compared to other cells.

The correlated background from cosmic muons is mainly dominated by the stopping muons (SM) in non-reactor environment. The selected SM arise from the muons entering through the top layers of PSBs in ISMRAN array, stopping inside the array, and eventually decaying into $e^{-}$ or $e^{+}$ with a large energy deposition in ISMRAN array~\cite{cosmic,danss_expt,cosmic_flux}. 

\section{Reconstruction of $\mathrm{E_{sum}}$ and $\mathrm{N_{bars}}$ for different radioactive sources in the ISMRAN array}
\begin{figure}[h]
\hspace{-2.9em}  
\includegraphics[width=18cm,height=6.2cm]{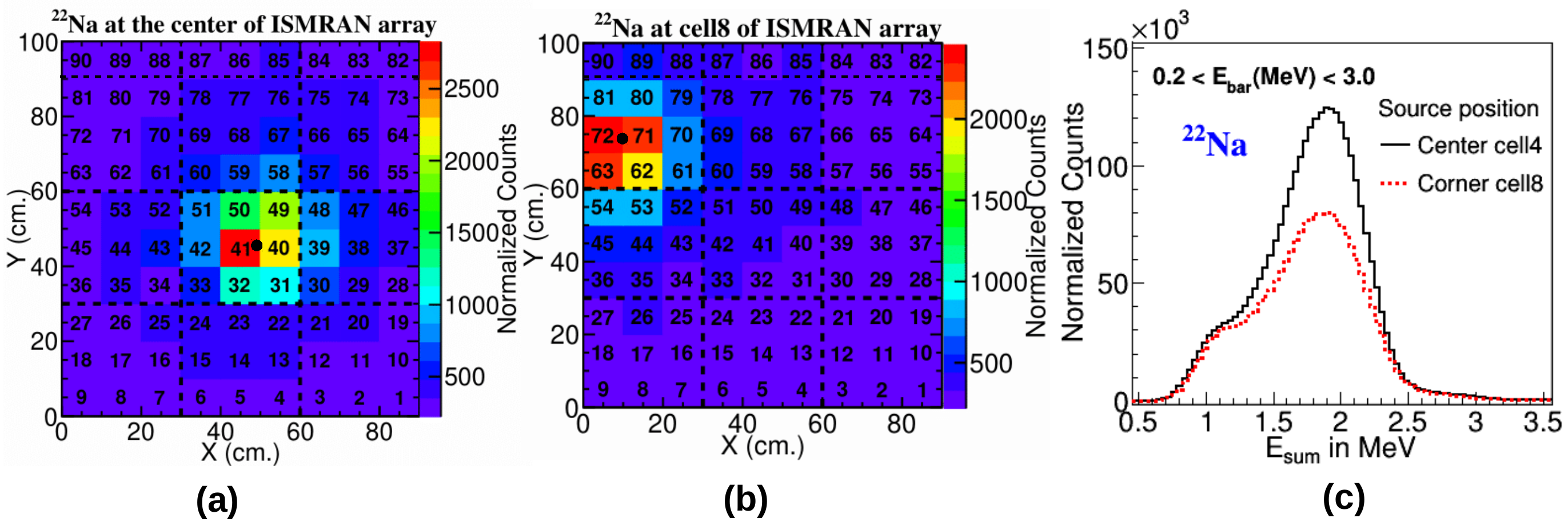}
\caption{Comparative study of the event distribution of $\gamma$-rays from $\mathrm{{}^{22}Na}$ source for the deployment of the source at panel (a) center cell, panel (b) corner cell. Also shown in panel (c) are the comparison of the $\mathrm{E_{sum}}$ distributions for the center and corner cell.}
\label{fig10}
\end{figure}

\begin{figure}[h]
\includegraphics[width=14.0cm,height=5.8cm]{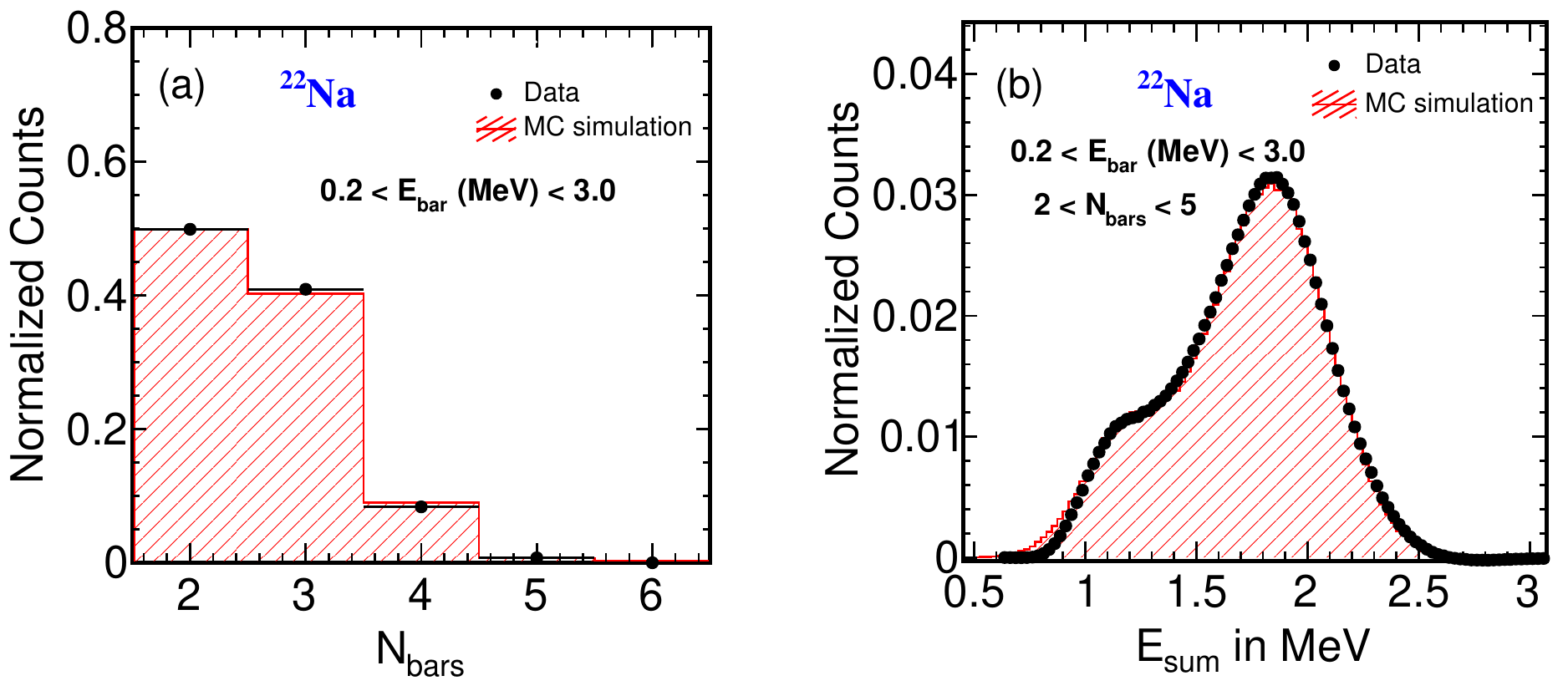}
\caption{ Panel (a) comparison of number of bars hit ($\mathrm{N_{bars}}$) distribution for $\mathrm{{}^{22}Na}$ source, placed at the center of ISMRAN array. Panel (b) comparison of measured and simulated sum energy ($\mathrm{E_{sum}}$) distribution for $\mathrm{{}^{22}Na}$ source, placed at the center of ISMRAN array.  }
\label{fig11}
\end{figure}
In order to understand the acceptance effect on the reconstruction of a coincident event, i.e. Compton scattered $\gamma$-ray energy deposition in multiple bars within a 20 ns coincidence time window, in ISMRAN setup, we use $\gamma$-rays + positron coming from $\mathrm{{}^{22}Na}$ radioactive source. The $\mathrm{{}^{22}Na}$ emits $\gamma$-rays of energy 0.511 MeV and 1.274 MeV. Figure ~\ref{fig10} (a) and (b) show the event distribution inside the ISMRAN array when the $\mathrm{{}^{22}Na}$ source is placed at the center cell 4 and the corner cell 8, respectively. As it can be seen qualitatively, the reconstructed events are contained within the vicinity of six bars from the central bar with the highest event counts in both cases. Figure~\ref{fig10} (c) shows the comparison of the $\mathrm{E_{sum}}$ distribution, in the range of $\mathrm{0.2 < E_{bar}~(MeV) < 3.0}$ and $\mathrm{2 < N_{bars} < 5}$, when the source is placed at the center cell 4 and corner cell 8. The correlated high energy $\gamma$-ray of 1.274 MeV and those from the annihilation of $e^{+}$ from $\mathrm{{}^{22}Na}$ source are reasonably well reconstructed at $\sim$ 2 MeV and at $\sim$ 1.0 MeV. The overall shift in the peak positions in both the cases is small and the acceptance of the higher energy $\gamma$-ray is better when the source is placed at the center cell 4 of ISMRAN as compared to the corner cell 8. This is due to the impartial containment of the higher energy $\gamma$-ray in the corner cell 8. At lower energy, the yields in both the cases are similar indicating the complete acceptance and reconstruction of annihilation $\gamma$-rays in the ISMRAN array. It is also important to compare the response of sum energy ($\mathrm{E_{sum}}$) and number of bars hit ($\mathrm{N_{bars}}$) distributions between data and MC simulations for optimizing the detection efficiencies using the selection criteria on these variables for the selection of prompt and delayed events for the candidate ${\overline{\ensuremath{\nu}}}_{e}$ events in data. Figure ~\ref{fig11} (a) and (b) show the comparison of reconstructed $\mathrm{N_{bars}}$ and $\mathrm{E_{sum}}$ distributions in data and MC simulation for $\mathrm{{}^{22}Na}$ source placed at the center cell 4 of the ISMRAN array. Only those PSBs are selected for the reconstruction of $\mathrm{N_{bars}}$ and $\mathrm{E_{sum}}$ where the individual energy deposition in each PSB is between 0.2 MeV to 3.0 MeV, so as to minimize the cosmic muon background. There is reasonably good agreement between data and MC results comparison for the $\mathrm{N_{bars}}$ and $\mathrm{E_{sum}}$ distributions.

The $\mathrm{N_{bars}}$ by a coincident event in the full ISMRAN array is a critical variable that affects the reconstructed sum energy of different radioactive $\gamma$ sources. Figure ~\ref{fig11}(a) shows the comparison of data and MC simulation of $\mathrm{N_{bars}}$ for reconstructing the sum energy of $\mathrm{{}^{22}Na}$ source. In earlier studies using same PSBs, it is observed that the ratio of signal from coincident events using $\mathrm{{}^{60}Co}$ radioactive $\gamma$-ray source to natural background events worsens with increase in $\mathrm{N_{bars}}$ in sum energy distribution inside a matrix of 4$\times$4 PSBs~\cite{ISMRAN}. Hence, by appropriately selecting $\mathrm{N_{bars}}$ and $\mathrm{E_{bar}}$ in each PSB, the coincident events can be reconstructed based on $\mathrm{E_{sum}}$ and $\mathrm{N_{bars}}$ variables for ISMRAN setup for coincident $\gamma$-rays generated in the decay of radioactive sources. As it can be seen from Fig ~\ref{fig11}(b), a peak at $\sim$1.78 MeV corresponds to the coincident of $\gamma$-ray and positron events from $\mathrm{{}^{22}Na}$ source and a feature at $\sim$1.2 MeV corresponds to the $\gamma$-ray of energy 1.274 MeV originating from $\mathrm{{}^{22}Na}$ source ~\cite{PROSPECT} in $\mathrm{E_{sum}}$ distribution, which has been reconstructed in the range of $\mathrm{2 < N_{bars} < 5}$. This method will be periodically tested in the real experiment at reactor hall while performing the in-situ calibrations using radioactive $\gamma$-ray sources for the uniformity of response in the PSBs.

\begin{figure}[h]
\begin{center}
\hspace{-2.9em}  
\includegraphics[width=15cm,height=6.2cm]{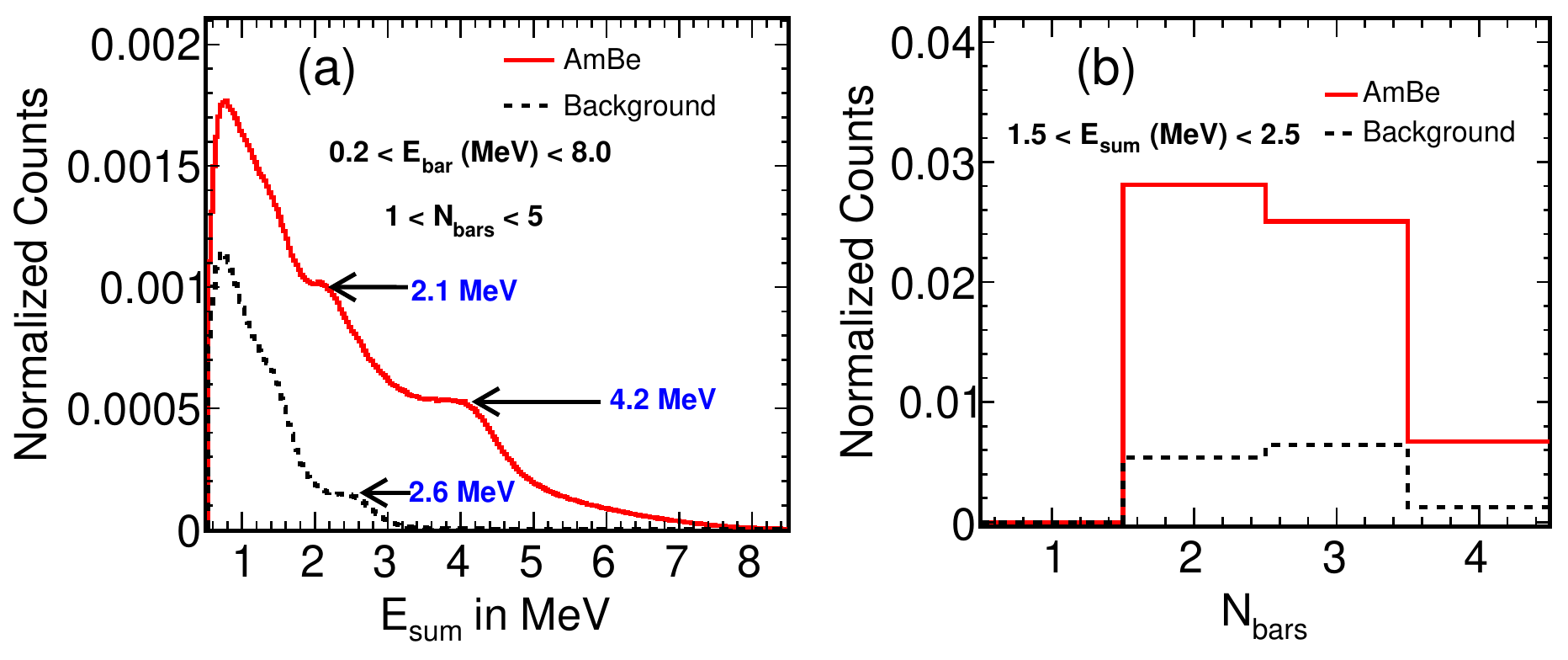}
\caption{ Panels (a) and (b) show the comparison of $\mathrm{E_{sum}}$ and $\mathrm{N_{bars}}$ distributions for n-H capture events from an Am/Be neutron source and accidental natural background events in ISMRAN array, respectively.}
\label{fig12}
\end{center}
\end{figure}

\begin{figure}[h]
\hspace{-2.9em}  
\includegraphics[width=17.5cm,height=5.6cm]{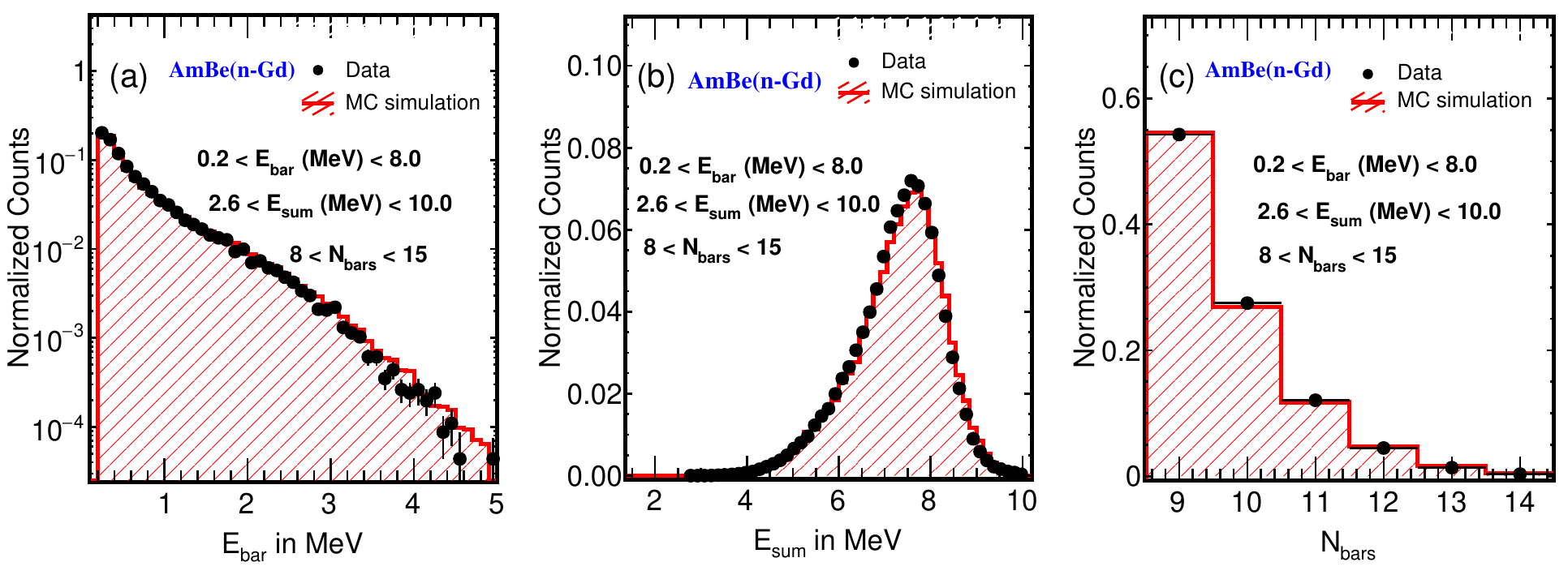}
\caption{Panels (a), (b) and (c) shows the comparison of measured data (solid) and MC simulated (shaded) individual $\mathrm{E_{bar}}$, $\mathrm{E_{sum}}$ and $\mathrm{N_{bars}}$ distributions for cascade $\gamma$-rays from n-Gd capture events for an Am/Be source placed on top of ISMRAN array, respectively.}
\label{fig13}
\end{figure}

\begin{figure}[h]
\begin{center}
\includegraphics[width=14cm,height=5.6cm]{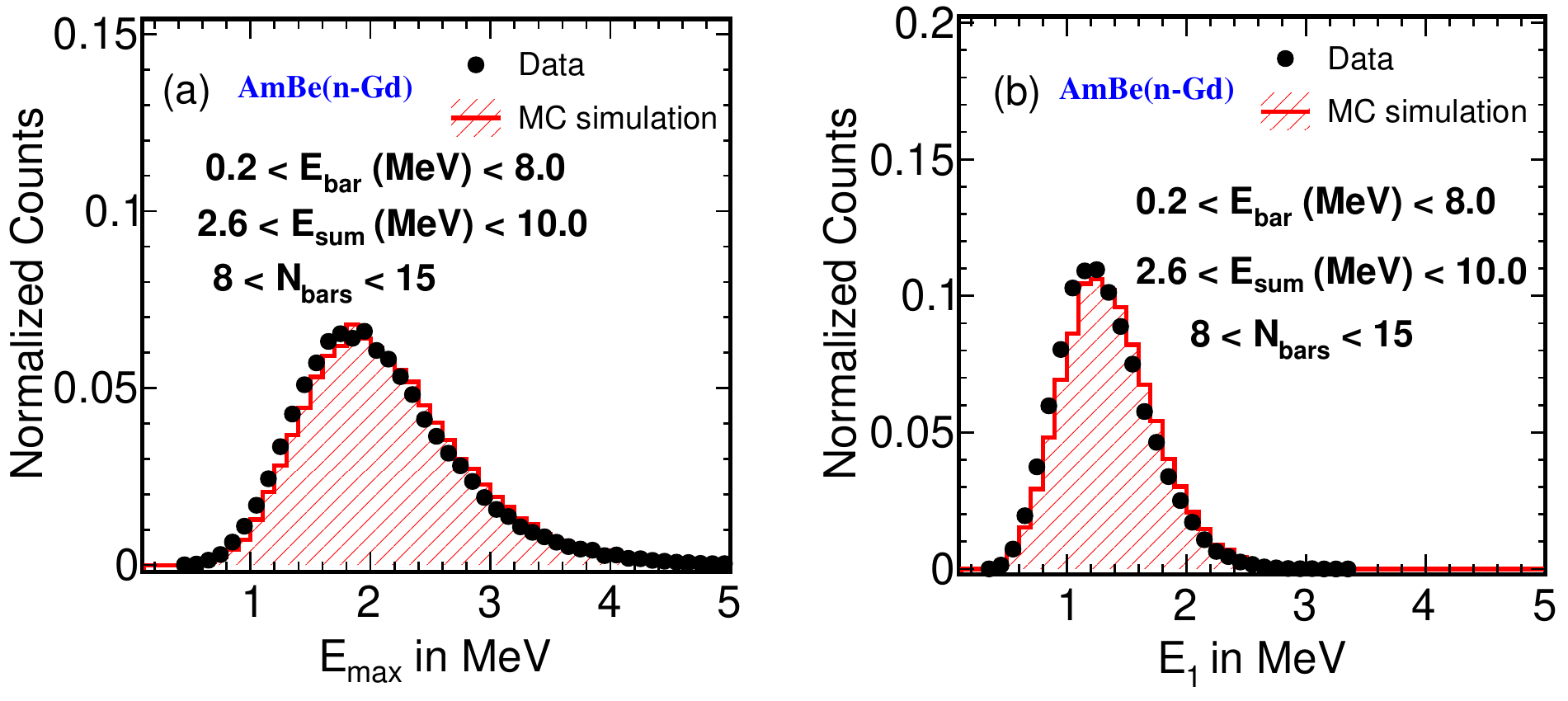}
\caption{Panels (a) and (b) show the comparison between measured data (solid) and MC simulated events (shaded) for $\mathrm{E_{max}}$ and $\mathrm{E_{1}}$ in  PSBs for cascade $\gamma$-rays from the n-Gd capture events using an Am/Be neutron source in ISMRAN array, respectively.}
\label{fig14}
\end{center}
\end{figure}

\begin{figure}
\begin{center}
\includegraphics[width=14cm,height=5.8cm]{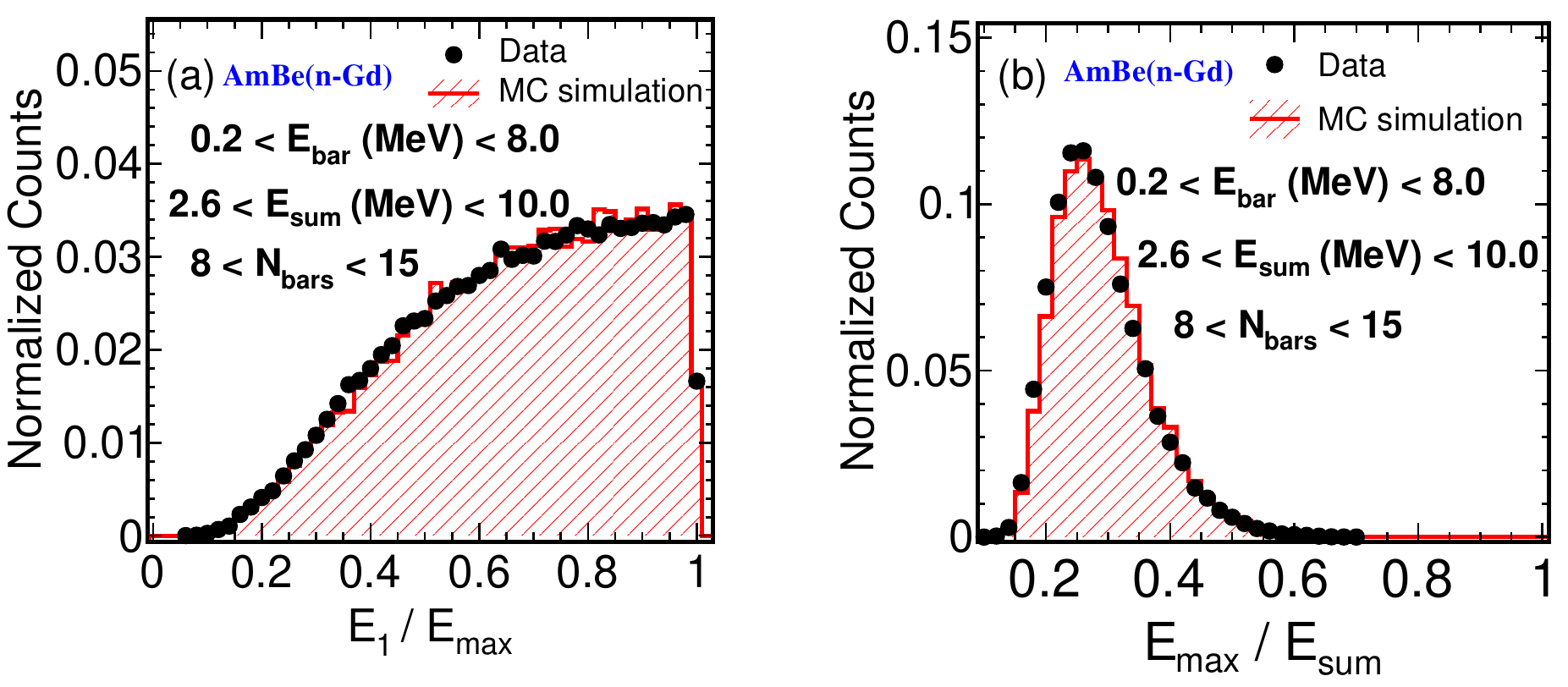}
\caption{Panel (a) and (b) show the comparison between measured data (solid) and MC simulated events (shaded) for energy ratios $\mathrm{E_{1}}$/$\mathrm{E_{max}}$ and $\mathrm{E_{max}}$/$\mathrm{E_{sum}}$ for cascade $\gamma$-rays from the n-Gd capture events using an Am/Be neutron source in ISMRAN array, respectively. }
\label{fig15}
\end{center}
\end{figure}

To have a good understanding of the ${\overline{\ensuremath{\nu}}}_{e}$ interaction, it is necessary to reconstruct the neutron capture event on Gd or H nuclei. The $\mathrm{E_{sum}}$ and $\mathrm{N_{bars}}$ hit variables have been reconstructed for n-H and n-Gd capture events using Am/Be neutron source. The Am/Be neutron source is placed on top of the ISMRAN array, since it was not possible to put the source at the center of the array due to the space constraints.
Figure~\ref{fig12} (a) shows the comparison of $\mathrm{E_{sum}}$  obtained from Am/Be neutron source and natural radioactive background. The $\mathrm{E_{sum}}$ is reconstructed using the individual PSBs satisfying the energy selection criteria between 0.2 MeV to 8.0 MeV and $\mathrm{2 < N_{bars} < 5}$. A peak at $\sim$ 2.1 MeV corresponds to the neutron capture on hydrogen and at $\sim$ 4.2 MeV corresponds to the 4.4 MeV high energy $\gamma$-ray originating from Am/Be source. Figure~\ref{fig12} (b) shows the comparison of $\mathrm{N_{bars}}$ between Am/Be neutron source and natural background. These selection criteria are made to compare the $\mathrm{E_{sum}}$ and $\mathrm{N_{bars}}$ variables between n-H capture events and natural background events. In both variables, n-H capture events are distinctively observed from the natural background.
The selection of slightly different $\mathrm{N_{bars}}$ cut for the n-H events from Am/Be source is to emphasis the ability of reconstruction of the mono-energetic $\gamma$-ray from n-H capture event in the presence of the underlying natural background events. For $\mathrm{{}^{22}Na}$ source, the underlying background events from the natural radioactivity are almost negligible and hence do not interfere in the sum energy spectra.
Figure~\ref{fig13} (a), (b) and (c) show the comparison between data and MC simulation for individual $\mathrm{E_{bar}}$, $\mathrm{E_{sum}}$ and $\mathrm{N_{bars}}$ from the cascade $\gamma$-rays from n-Gd capture events using an Am/Be neutron source, respectively. Only those PSBs are selected for the $\mathrm{E_{sum}}$ where the individual energy deposition in each PSB is between 0.2 MeV to 8.0 MeV. For the complete capture and reconstruction of cascade $\gamma$-rays, with $\mathrm{E_{sum}}$ $\sim$ 8 MeV, in the ISMRAN array, the $\mathrm{E_{sum}}$ is required to be in the range of 2.6 MeV to 10.0 MeV and $\mathrm{N_{bars}}$ should be in the range of 9 to 14. This selection criteria is implemented to reduce the contamination from the natural radioactive and cosmogenic muon background in the  $\mathrm{E_{sum}}$. The cascade $\gamma$-rays from the n-Gd neutron capture events were implemented using the DICEBOX packages in GEANT4 separately and then propagated for the recording of the energy deposition in each PSB. The energy resolution function is incorporated, for the deposited energy in each PSB, in the MC simulated events. A good agreement between data and MC simulation has been observed in the individual cascade $\gamma$-ray distribution, as shown in Fig~\ref{fig13} (a), for n-Gd capture events. Figure~\ref{fig13} (b) shows the comparison of the reconstructed $\mathrm{E_{sum}}$ from cascade $\gamma$-rays between data and MC simulated events. A clear peak at $\sim$ 7.7 MeV is observed in data and is in reasonable agreement with that obtained from MC simulated events. This again shows the uniformity and linearity in the energy response across the PSBs in ISMRAN detector. Figure~\ref{fig13} (c) shows the comparison of the reconstructed $\mathrm{N_{bars}}$ distribution between data with MC simulated events for the cascade $\gamma$-rays from n-Gd capture events. Again, a reasonably good agreement between data and MC simulated events is observed and a selection criteria on $\mathrm{N_{bars}}$ can filter out the complete containment of the cascade $\gamma$-ray event from n-Gd capture in the ISMRAN array.
Additional criteria for selecting the cascade $\gamma$-rays of n-Gd capture events from natural background, we sort the hit bars in each reconstructed event, according to the decreasing order of their individual energy depositions in PSBs. The PSB which has maximum energy deposition, is denoted as $\mathrm{E_{max}}$, out of all the bars used for the reconstruction of the candidate events. Similarly $\mathrm{E_{1}}$ correspond to the bar which is next, in terms of energy deposition with respect to the $\mathrm{E_{max}}$, for the reconstruction of the candidate event. Figure~\ref{fig14} (a) and (b) show the comparison of reconstructed $\mathrm{E_{max}}$ and $\mathrm{E_{1}}$ distributions in data compared with MC simulation for the $\mathrm{E_{sum}}$ in the range of 2.6 MeV to 10.0 MeV and the $\mathrm{N_{bars}}$ within the range of 9 to 14. A reasonable agreement between data and MC simulated events for the $\mathrm{E_{max}}$ and $\mathrm{E_{1}}$ distributions are observed with average $\chi^{2}$ is 0.239. More differential comparison of the energy profiles in PSB is done using the ratio of $\mathrm{E_{1} / E_{max}}$ and $\mathrm{E_{max} / E_{sum}}$ for the n-Gd capture events. Figure ~\ref{fig15} (a) and (b) show the comparison of reconstructed $\mathrm{E_{1} / E_{max}}$ and $\mathrm{E_{max} / E_{sum}}$ distributions in data compared with MC simulated events for cascade $\gamma$-rays from n-Gd capture events. A reasonable agreement between data and MC simulated events for the $\mathrm{E_{1} / E_{max}}$ and $\mathrm{E_{max} / E_{sum}}$ distributions are observed with average $\chi^{2}$ is 0.179. Applying selection criteria on the $\mathrm{E_{1} / E_{max}}$ and $\mathrm{E_{max} / E_{sum}}$ ratios will improve the rejection of fast neutron or accidental background from ${\overline{\ensuremath{\nu}}}_{e}$ delayed events~\cite{Roni}.
A detailed study of the peak positions and the resolution of the peaks using reconstructed $\mathrm{E_{sum}}$ obtained from different radioactive and neutron source are tabulated in table~\ref{table2}. A very good agreement is observed between the reconstructed $\mathrm{E_{sum}}$ from data and MC simulated events.
\begin{table}[h]
\begin{small}
  \begin{center}
  \caption{Comparison of the Compton edges (CE) $\mathrm{E_{sum}}$ between data and MC results for different radioactive sources. Peak energies and $\sigma$ are given in MeV. Statistical errors of the fit results are negligible.}
  \label{table2}
\begin{tabular}{|c|c|c|c|c|c|}
\hline
\makecell{Radioactive sources} & $\mathrm{E_{data}}$ (MeV) & $\mathrm{E_{mc}}$ (MeV) & $\mathrm{\sigma_{data}}$ (MeV) & $\mathrm{\sigma_{mc}}$ (MeV) & $\mathrm{{\chi}^2}$  \\
\hline
\makecell{$\mathrm{{}^{22}Na}$} & 1.78 & 1.79 & 0.27 & 0.29 & 1.033\\
\hline
\makecell{Am/Be:H(n, $\gamma$ )}        & 2.10 & 2.15 & 0.66 & 0.30 & -\\
\hline
\makecell{Am/Be:Gd(n, $\gamma$ )}       & 7.70 & 7.65 & 0.73 & 0.78 & 0.258 \\
\hline
\end{tabular}
\end{center}
\end{small}
\end{table}  

\section{Study of fast neutron energy response with the ISMRAN array using TOF technique}
\begin{figure}[h]
\begin{center}
\hspace{-3.2em}  
\includegraphics[width=14cm,height=8cm]{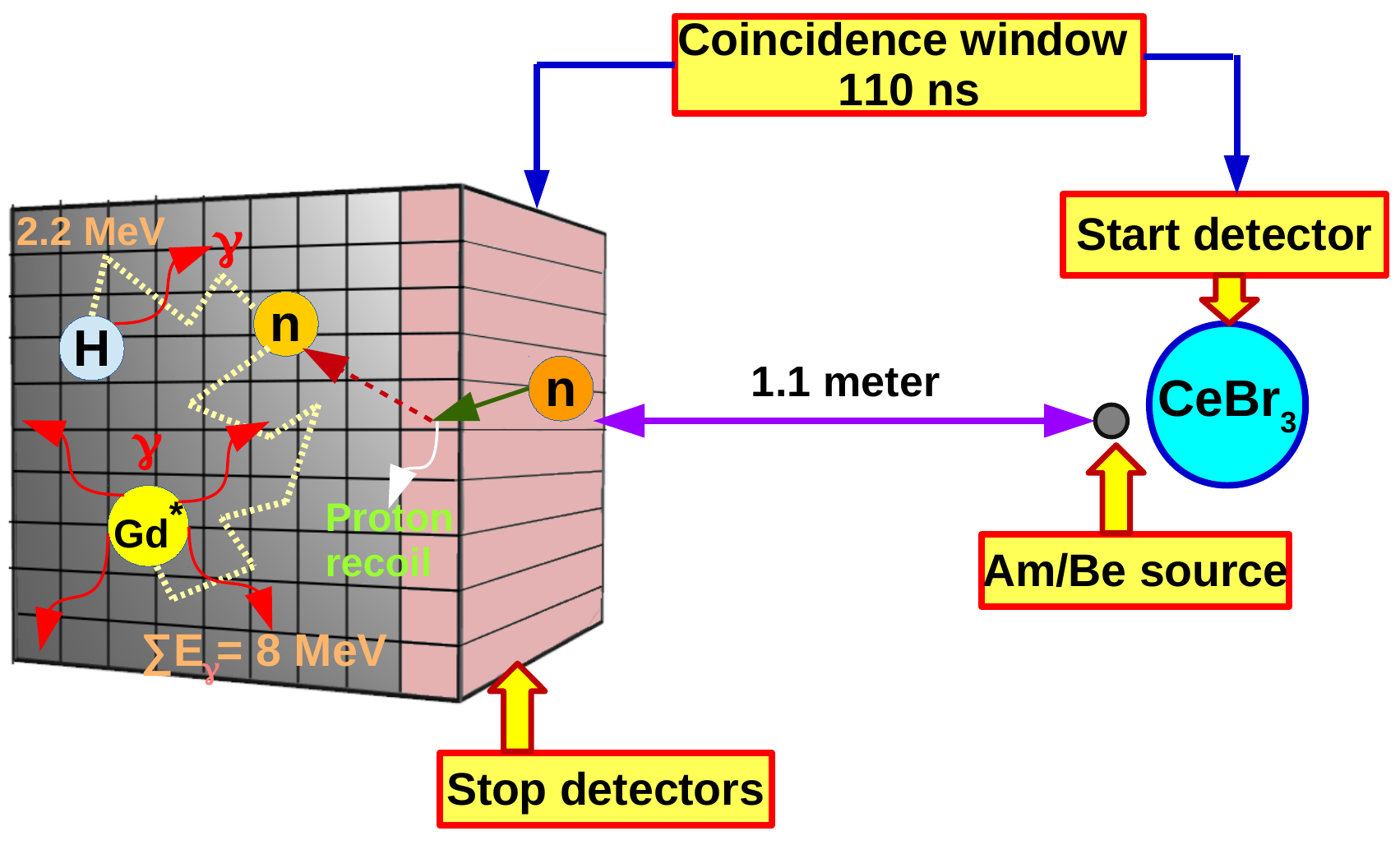}
\caption{The schematic representation of TOF experimental setup, illustrating the fast neutron event in ISMRAN detector mimicking the prompt and delayed event signatures of IBD.}
\label{fig16}
\end{center}
\end{figure}

Fast neutron (FN) is an important source of correlated background for ISMRAN detector setup, apart from the cosmogenic and natural radioactive backgrounds. ISMRAN deals with an active FN background inside the reactor hall, predominantly generated by the spontaneous fission of the $(\mathrm{{}^{238}U})$ and $(\mathrm{{}^{234}Th})$ decay chains from reactor fuel and $(\alpha,n)$ reactions in surrounding material inside the reactor hall. FNs are also produced by cosmic muons in the inactive (shielding) regions surrounding the detector~\cite{reno_tof1}. Their large interaction length allows them to cross the detector, get thermalized and then get captured inside the ISMRAN array, causing both a prompt-like event triggered by recoil protons and a delayed-like event triggered by capture on Gd or H nuclei, which mimics exact signature of a prompt and a delayed pair from IBD process as shown in Fig.~\ref{fig16}. Reactor related fast neutrons typically have energies of few MeV and neutrons of much higher energies (up to a few GeV) are produced from cosmogenic neutrons and cosmic muon-induced spallation reactions. Therefore, a good understanding of fast neutron response in PSBs is an essential pre-requisite for suppression and discrimination of fast neutron background from IBD events~\cite{reno_tof2}.
Fast neutron energy response has been measured in PSB, using time-of-flight (TOF) technique, for neutrons of energy between 1 MeV to 8 MeV originating from an Am/Be source. Fast neutrons are produced via $\mathrm{{}^{9}Be}(\alpha, n)\mathrm{{}^{12}C}$ reactions, where the $\alpha$ particle is produced in radioactive decays of $\mathrm{{}^{241}Am}$. In about $\sim$60$\%$ of the cases, the carbon nucleus is produced in an excited state, and emits a $\sim$ 4.44 MeV $\gamma$-ray in addition to the neutron.
To get background-free fast neutron events, the 4.44 MeV $\gamma$-ray is used as a tag for the delayed coincidence measurement of the fast neutrons using TOF technique. We used a 2$''$ cerium bromide ($\mathrm{CeBr_{3}}$) detector for triggering 4.44 MeV $\gamma$-ray, which provides the reference start time for the corresponding emitted neutrons. The $\mathrm{CeBr_{3}}$ detector is placed very close to the Am/Be source. The ISMRAN array is positioned at $\sim$1.1 m away from the Am/Be source at the source height, as shown in Fig.~\ref{fig16}. As a trigger for the start time, 4.44 MeV $\gamma$-ray has been tagged in $\mathrm{CeBr_{3}}$ detector and neutron or $\gamma$-ray as stopped time candidate is recorded at the first column (10 PSBs) of ISMRAN array. The time coincidence window between start and stop detector is chosen to be 110 ns. The schematic representation of the entire setup for this technique is shown in Fig.~\ref{fig16}. By recording the start and stop time signals from $\mathrm{CeBr_{3}}$ and first column of ISMRAN array, the TOF is reconstructed for discrimination between the $\gamma$-rays and neutrons on the first column of the ISMRAN array. The time difference between the neutron on the first column and the $\mathrm{CeBr_{3}}$ detector, a neutron time-of-flight (TOF) is constructed and the kinetic energy of the fast neutrons are determined.
One of the most important variable in identifying ${\overline{\ensuremath{\nu}}}_{e}$ candidate events is the time difference between prompt and delayed events inside the ISMRAN array. From MC simulations, the average time difference between prompt and delayed event is observed to be $\sim$ 68 $\mathrm{\mu}s$ with a range extending up to 250 $\mathrm{\mu}s$~\cite{ISMRAN}. Using these tagged fast neutrons in first column of the ISMRAN array and subsequently using them as triggers, the neutron capture time can be measured in the rest of the ISMRAN array. 
\begin{figure}[h]
\begin{center}
\hspace{-2.9em}  
\includegraphics[width=15cm,height=6cm]{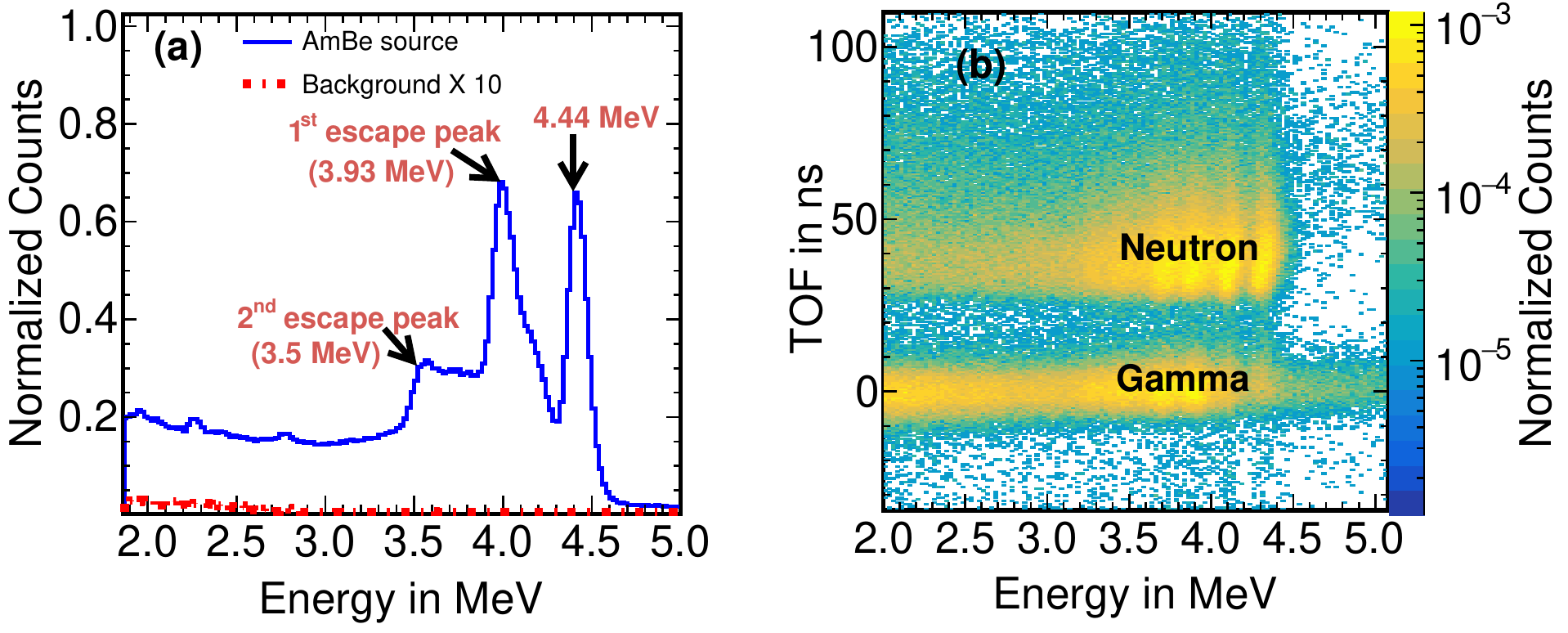}
\caption{ Panel (a) shows the comparison of the $\gamma$-ray energy spectrum obtained from Am/Be source and the natural radioactive background in $\mathrm{CeBr_{3}}$ detector. The blue histogram corresponds to the $\gamma$-rays measured from the Am/Be source and the red histogram corresponds to the $\gamma$-rays measured without source. Panel (b) shows the time-of-flight (TOF), measured in the first column in PSBs in ISMRAN array, as a function of the $\gamma$-ray energy in $\mathrm{CeBr_{3}}$ detector within the time coincidence of 110 ns between PSBs and $\mathrm{CeBr_{3}}$ detector.}
\label{fig17}
\end{center}
\end{figure}
Figure~\ref{fig17} (a) shows the comparison of energy deposition of the $\gamma$-rays from an Am/Be neutron source and the natural background in $\mathrm{CeBr_{3}}$ detector. The $\mathrm{CeBr_{3}}$ detector is calibrated using standard radioactive $\gamma$-ray sources and the energy resolution obtained at 0.662 MeV is 3.8$\%$. An energy threshold is applied at around 2 MeV to exclude very low energy $\gamma$-ray events, to avoid the accidental random correlations. The features in the $\gamma$-ray distribution from Am/Be between 3 MeV to 5 MeV correspond to neutron-associated $\gamma$-rays from the de-excitation of $\mathrm{{}^{12}C^{*}}$. As it can be seen from Fig~\ref{fig17}(a), a full-energy peak at $\sim$4.44 MeV corresponds to high energy $\gamma$-ray due to the de-excitation of carbon nucleus from Am/Be neutron source and while the corresponding first and second escape peaks appear at $\sim$3.93 MeV and $\sim$3.5 MeV, respectively. The natural $\gamma$-ray background is scaled by a factor of 10 for the representation purposes and is negligible in the current measurements. Figure ~\ref{fig17} (b) shows the TOF distribution of fast neutrons as a function of $\gamma$-ray energy deposition in $\mathrm{CeBr_{3}}$ detector within the time coincidence window of 110 ns. Two distinct bands at $\sim$1.0 ns and $\sim$ 38 ns in the TOF distribution are due to $\gamma$-rays and neutron, respectively. The separation between $\gamma$-rays and neutrons is excellent and this feature is used to tag the fast neutrons on the first column of the ISMRAN array. 
\begin{figure}[h]
\begin{center}
\hspace{-2.9em}  
\includegraphics[width=14.5cm,height=6cm]{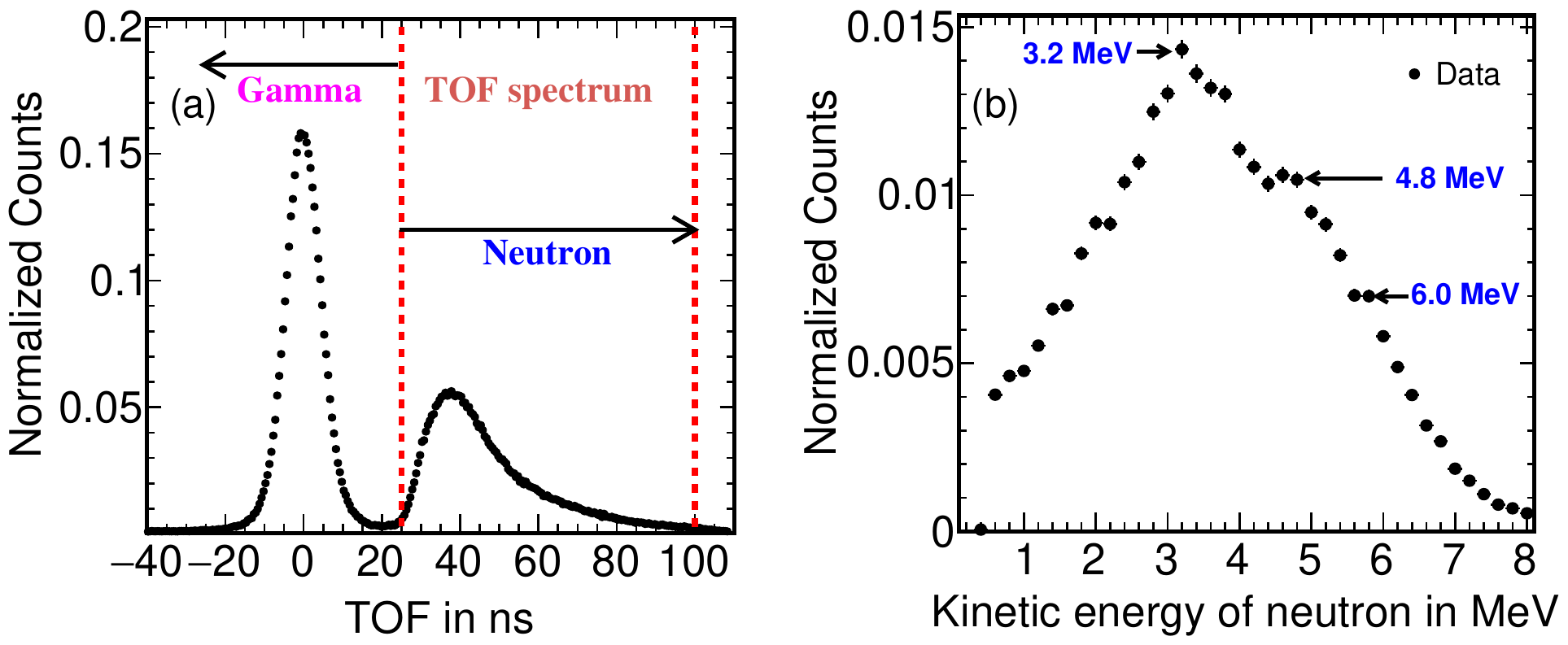}
\caption{Panel (a) shows the projection of TOF distribution for $\gamma$-rays and fast neutrons. Panel (b) shows the derived kinetic energy distribution of fast neutron from an Am/Be source using TOF technique.}
\label{fig18}
\end{center}
\end{figure}
Figure~\ref{fig18} (a) shows the projection of TOF distribution for $\gamma$-rays and fast neutrons by tagging high energy $\gamma$-rays between 3.0 MeV to 5.0 MeV in $\mathrm{CeBr_{3}}$ detector. A peak at $\sim 1.0$ ns correspond to the $\gamma$-rays, having width of $\sim$4.5 ns. $\mathrm{T_{0}}$, the instant of emission of the neutron from the source, is determined from the location of this peak in the TOF spectra using the speed of light (c) and the measurement of the distances (L) between the PSBs and the neutron source. The $\gamma$-rays and the fast-neutron are clearly identified and separated in the TOF distribution. The tagged neutron peak is around $\sim 38$ ns and ranges between 25 ns to 100 ns. Events above TOF $>$ 100 ns are mainly due to the random coincidences. The kinetic energy distribution of the neutrons can be determined by measured TOF using the following classical expression of neutron kinetic energy, 

\begin{equation}\label{eq:tof1}
\mathrm{ E_{n}} = \mathrm{\frac{1}{2}mV^{2}} = \mathrm{\frac{1}{2}m \left(\frac{L^{2}}{t^{2}}\right)} = \mathrm{\alpha^{2}\left(\frac{L^{2}}{t^{2}}\right)}; \mathrm{where,}  \mathrm{\alpha = 72.3(\sqrt{ev}).\mu s/m},
\end{equation}
\begin{equation}\label{eq:tof2}
\mathrm{TOF_{n}} = \mathrm{\frac{72.3 L}{\sqrt{E_{n}}}};
\mathrm{TOF_{\gamma}} = \left(\frac{L}{c}\right).
\end{equation}
Figure~\ref{fig18} (b) shows the neutron kinetic energy distribution, derived from the TOF spectrum of tagged neutron between 25 ns and 100 ns. The neutron kinetic energy distribution shows peaks at 3.2 MeV, 4.8 MeV and 6.0 MeV, which is in good agreement with Lorch data~\cite{Lorch} between 2.5 MeV and 8 MeV.
Figure ~\ref{fig19} (a) shows the energy deposition ($\mathrm{E_{dep}^{p}}$) of the recoiling proton by fast neutrons, in $\mathrm{MeV_{ee}}$, in PSB as a function of kinetic energy of neutron derived from TOF. To minimize the contributions from the accidental natural background, $\gamma$-ray energy selection between 3.0 MeV and 5.0 MeV is made which covers the first escape, second escape and photoelectric peak for 4.44 MeV $\gamma$-ray from Am/Be source. For reducing the uncertainties while deriving the kinetic energy of the neutrons, we select the z-position of the stopped events within $\pm$ 10 cm around the center of the PSBs. The projection of the $\mathrm{E_{dep}^{p}}$ of the neutron in PSB for different neutron kinetic energy bins is shown in Fig ~\ref{fig19} (b). The width of the projected $\mathrm{E_{dep}^{p}}$ distribution increases with increasing kinetic energy of the neutron. This is due to the fact that exact binning in the TOF distribution for deriving kinetic energy is not possible and the smearing effect on kinetic energy is observed more prominently towards smaller TOF values which yields larger neutron kinetic energies. To reduce this smearing effect, the projection of $\mathrm{E_{dep}^{p}}$ by the fast neutrons in PSBs are plotted in bins of 1 MeV for the derived kinetic energy of the fast neutrons. Also for fast neutrons with higher energies, the multiple scattering within PSBs can be characterized by broader signal pulses because of their successive convolution of individual pulses separated in time. The peak values of the projected distribution of $\mathrm{E_{dep}^{p}}$ by the fast neutrons through recoiling proton are used for the parametrization of the deposited energy as a function of neutron kinetic energy in PSBs. This parametrization would help in determining the fast neutron energy scale from reactor background in the ISMRAN array.
\begin{figure}[h]
\begin{center}
\hspace{-2.9em}  
\includegraphics[width=14.4cm,height=5.8cm]{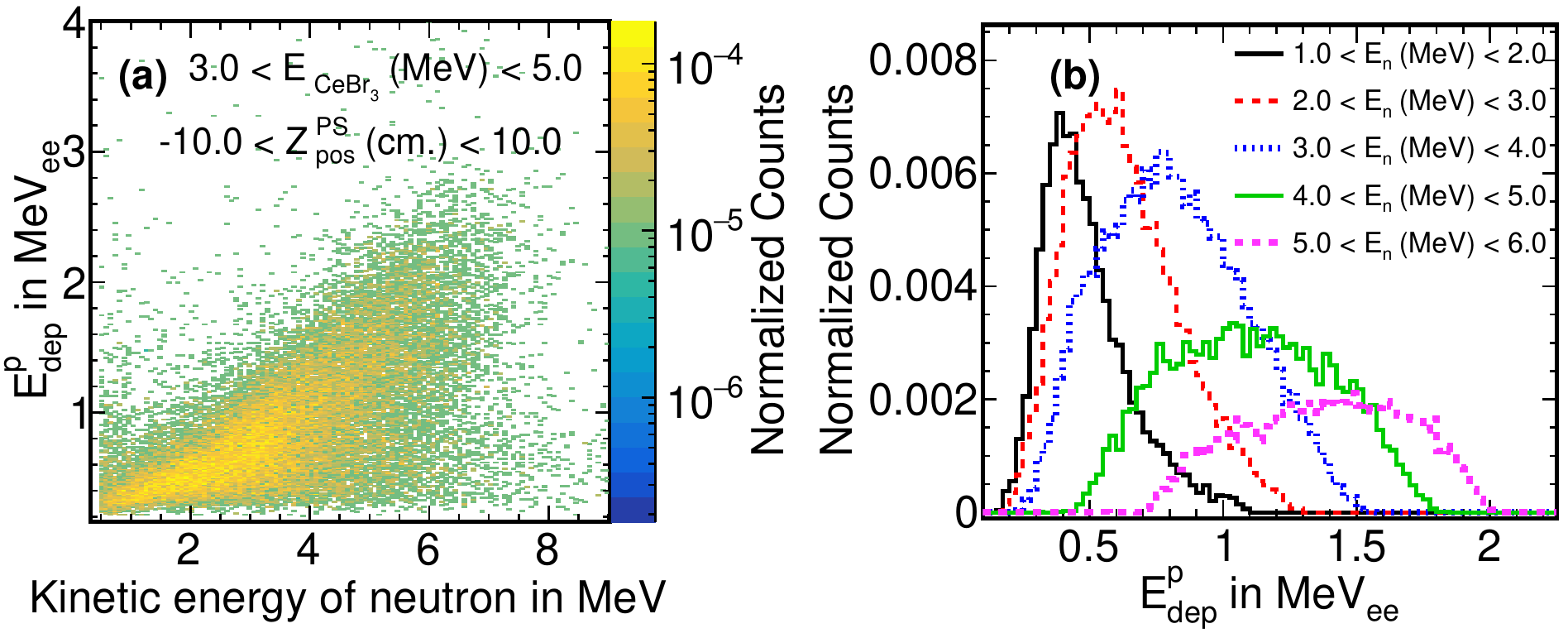}
\caption{Panel (a) shows the $\mathrm{E_{dep}^{p}}$ of tagged neutron in PSBs as a function of kinetic energy of tagged neutron and panel (b) shows the projected distribution of $\mathrm{E_{dep}^{p}}$ of neutron for different kinetic energy bins of tagged neutron.}
\label{fig19}
\end{center}
\end{figure}
 Figure~\ref{fig20} shows the mean of the $\mathrm{E_{dep}^{p}}$, obtained from Fig~\ref{fig19} (b), of the fast neutrons in PSB as a function of kinetic energy of neutrons. To obtain a relation between the $\mathrm{E_{dep}^{p}}$ by the recoiling protons from fast neutrons in PSB,  a parametrization between kinetic energy of neutron and $\mathrm{E_{dep}^{p}}$ from recoiling proton in PSB is obtained using the empirical formula as shown in Eqs.~\ref{eq:tof_formula}
\begin{equation}\label{eq:tof_formula}
  \mathrm{E_{dep}^{p}} = \mathrm{a_{0}E_{n} - a_{1}(1 - e^{ -a_{2}E_{n}^{a_{3}}}) }
\end{equation}
\begin{figure}[h]
\begin{center}
\includegraphics[width=6.5cm,height=5.7cm]{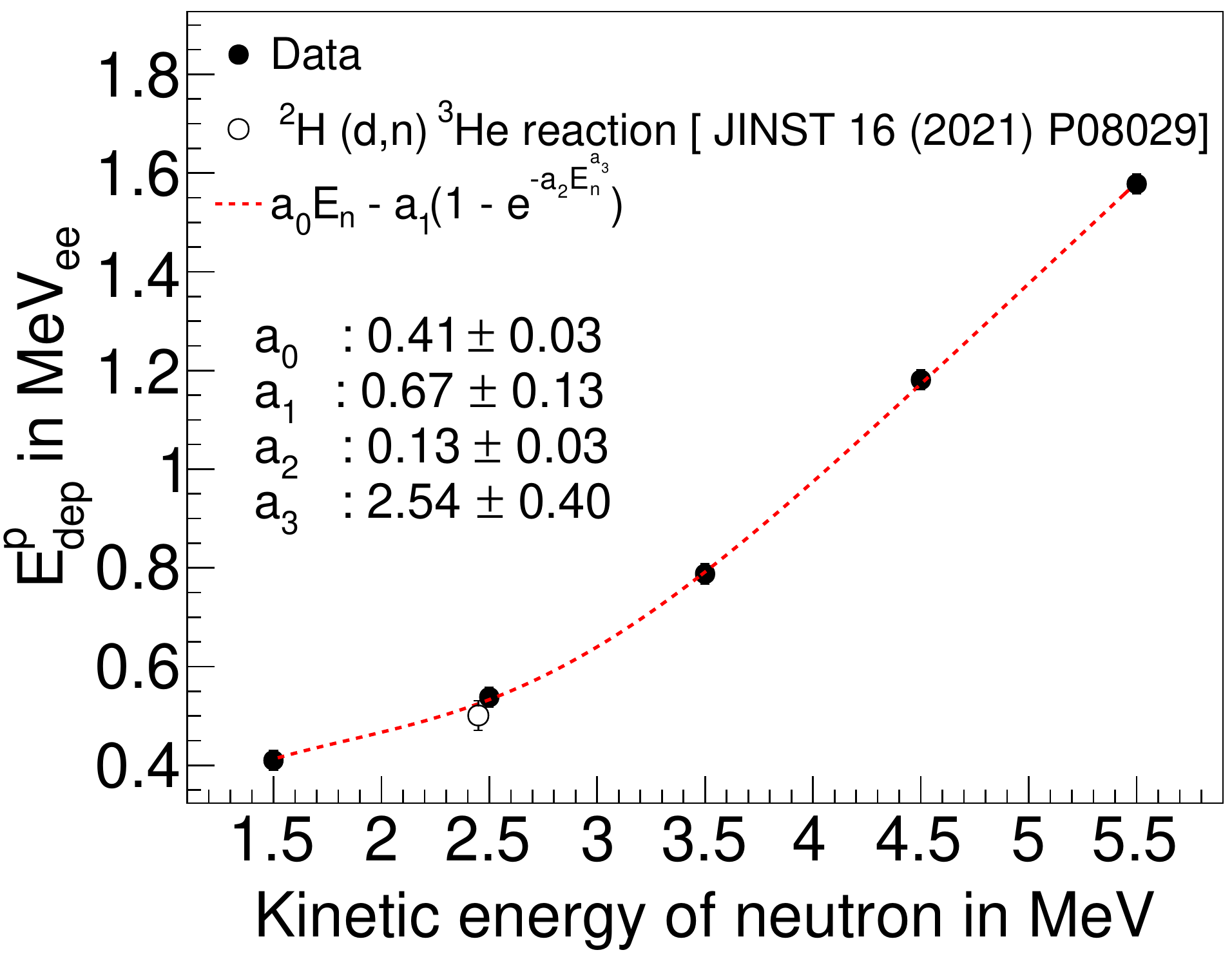}
\caption{Parametrization of $\mathrm{E_{dep}^{p}}$ due to proton recoils as a function of neutron kinetic energy in PSB.}
\label{fig20}
\end{center}
\end{figure}
where, $\mathrm{E_{dep}^{p}}$ is the electron energy equivalent of deposited energy due to proton recoil and $\mathrm{E_{n}}$ is the kinetic energy of fast neutron. The parameters $\mathrm{a_{0}, a_{1}, a_{2}, and~a_{3}}$ are represent the light response function for fast neutron in PSB ~\cite{tof_para}.

\begin{figure}[h]
\hspace{-3.9em}  
\includegraphics[width=18.5cm,height=5.4cm]{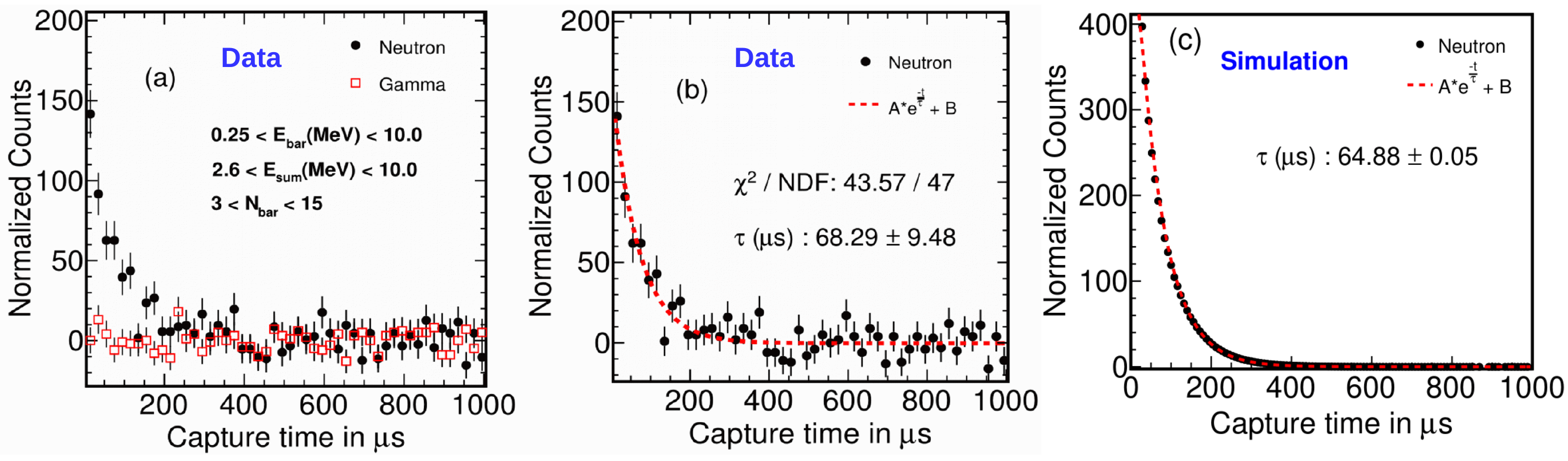}
\caption{ Panel (a) shows the measured $\Delta T$ distributions for the start obtained from the first column of the ISMRAN using the tagged fast neutron or $\gamma$-ray events with stop recorded in the rest of the ISMRAN array. Panel (b) shows the fit result for $\Delta T$ distribution between tagged fast neutron from first column in ISMRAN array and stop from rest of the ISMRAN array. Panel (c) shows the characteristic time difference, obtained from MC simulations, for the n-Gd capture events for fast neutron from Am/Be source in ISMRAN array.}
\label{fig21}
\end{figure}
The measurement of the characteristic time difference of the neutron thermalization and capture on Gd nucleus is one of the most important variables used in the discrimination of ${\overline{\ensuremath{\nu}}}_{e}$ candidate events. With these tagged neutrons obtained from TOF technique at the first column of the ISMRAN array, we demonstrated a novel technique to measure the characteristic capture time of these neutrons on Gd nucleus. We use the tagged fast neutrons identified in the first column of ISMRAN array using TOF technique, treating them as prompt events, and then searching for the n-Gd capture event as a delayed candidate events in rest of the ISMRAN array within a coincidence time window of 1000 $\mu$s. Only those PSBs are selected for the $\mathrm{E_{sum}}$ of delayed candidate events where the individual $\mathrm{E_{bar}}$ in each PSB is between 0.25 MeV to 10.0 MeV. For delayed event candidate, the $\mathrm{E_{sum}}$, is required to be in the range of 2.6 MeV to 10.0 MeV and $\mathrm{N_{bars}}$ should be in the range of 4 to 14. Figure~\ref{fig21} (a) shows the time difference $\Delta T$ distribution between the start event time from the tagged neutrons and $\gamma$-ray events in the first column of the ISMRAN array and the stopped event time measured in the rest of the ISMRAN array.
The $\Delta T$ distribution for the neutron tagged events, shown in black solid points, shows a exponential behaviour which is the characteristic nature of the n-Gd capture events in ISMRAN array. One the other hand, the $\Delta T$ distribution for $\gamma$-ray tagged events, shown in red squares, shows a uniform distribution in $\Delta T$ indicating the randomness in the prompt-delayed event pairs. Figure~\ref{fig21} (b) shows the characteristic time $\tau$ obtained from the a fit with a exponential and a constant term. The exponential part represent the neutron capture time on Gd and a constant term, B, shows the accidental background. The $\tau$ obtained from the fit to the neutron tagged events from data is 68.29  $\pm$ 9.48 $\mu$s and is in good agreement with 64.88 $\pm$ 0.05 $\mu$s that obtained from MC simulation events as shown in Fig.~\ref{fig21} (c). This way we have demonstrated a novel technique for the determination of the neutron capture time on Gd in ISMRAN array inspired by the data driven method.
\section{Conclusions and Outlook}
We present the measurements of cosmogenic, natural radioactive backgrounds and detailed characterization of full scale ISMRAN array consisting of 90 plastic scintillator bars, at DIL, BARC. Description of the optical transmission, energy resolution and energy non-linearity is obtained for the PSBs for different radioactive $\gamma$ sources and are compared with GEANT4 based MC simulations. An energy resolution of $\sim$14$\%$ at 1 MeV has been achieved for the PSBs. Correlated and uncorrelated non-reactor backgrounds mainly coming from the natural radioactive $\gamma$-ray and cosmogenic muon background, are discussed in detail, which will be useful for discrimination of background events from the true IBD events. Sum energy and number of bars hit variables have been validated by comparing with GEANT4 based MC simulations for radioactive $\gamma$-ray + positron source such as $\mathrm{{}^{22}Na}$ placed at the center of the ISMRAN array. Sum energy, number of bars hit, energy deposition ratios variables have been compared with GEANT4 MC simulations for better understanding of the delayed events from neutron capture on Gd nucleus in ISMRAN array. The fast neutron energy response in ISMRAN detector is studied with an Am/Be source using TOF technique. This technique enabled the mapping of the response of the PSB to fast neutrons as a function of their kinetic energy, which is useful to get the energy deposition due to recoiling protons in PSBs. Using a novel technique of tagging fast neutron to measure the mean capture time of neutron in ISMRAN array is demonstrated in detail. The measured neutron capture time is 68.29 $\pm$ 9.48 $\mu$s and is in good agreement with the MC simulation results. Discrimination of fast neutron background from the prompt IBD events can be achieved by using the segmented geometry of ISMRAN array and combining energy dependent variable such as energy deposition ratios with other topological event selection cuts in PSBs along with the implementation of an advanced machine learning algorithms.

The full scale ISMRAN experiment has been installed and commissioned in the Dhruva reactor hall on a movable base structure. Arrangement of the passive shielding of 10 cm of lead and 10 cm of borated polyethylene are mounted on the base structure. The physics data campaign has started at the end of year 2021.

\section{Acknowledgments}
We are thankful to EmA$\&$ID workshop for building the temporary structure for placing the ISMRAN detectors in DIL. We would like to thank Dr. U.~K.~Pal, NPD BARC for providing us the Cerium Bromide detector for the background measurements and time of flight experiment. We would also like to thank Dr. D. Mulmule for his support while assembling the ISMRAN setup at DIL, BARC. 
\bibliographystyle{unsrt}  


\begin{thebibliography}{100}


\bibitem{Cowan}
  C.~L.~Cowan, F.~Reines, F.~B.~Harrison, H.~W.~Kruse, and A.~D.~McGuire,
  \newblock Detection of the Free Neutrino: a Confirmation.
  \newblock {\em Science}, 124, 103 (1956) .
  \newblock \url{https://www.science.org/doi/10.1126/science.124.3212.103}.

\bibitem{PDG}
  M.~Tanabashi, et~al.,
  \newblock Review of Particle Physics, (Particle Data Group),
  \newblock {\em Phys. Rev. D}, 98 (2018) 030001.
  \newblock \url{https://journals.aps.org/prd/abstract/10.1103/PhysRevD.98.030001}.

\bibitem{DayaBay}
  F.~P.~An, et~al.,
  \newblock Observation of Electron-Antineutrino Disappearance at Daya Bay.
  \newblock {\em Phys. Rev. Lett.}, 108 (2012) 171803.
  \newblock \url{https://link.aps.org/doi/10.1103/PhysRevLett.108.171803}.
  
\bibitem{RENO}
  J.~K.~Ahn, et~al., (RENO Collaboration),
  \newblock Observation of Reactor Electron Antineutrinos Disappearance in the RENO Experiment.
  \newblock {\em Phys. Rev. Lett.}, 108 (2012) 191802.
  \newblock \url{https://link.aps.org/doi/10.1103/PhysRevLett.108.191802}.
  
\bibitem{DChooz}
  Y.~Abe, et~al., (Double Chooz Collaboration).
  \newblock Indication of Reactor ${\overline{\ensuremath{\nu}}}_{e}$ Disappearance in the Double Chooz Experiment.
  \newblock {\em Phys. Rev. Lett.}, 108 (2012) 131801.
  \newblock \url{https://link.aps.org/doi/10.1103/PhysRevLett.108.131801}.
  
\bibitem{DBay}
  D.~Adey et al., (The Daya Bay Collaboration).
  \newblock Measurement of the Electron Antineutrino Oscillation with 1958 Days of Operation at Daya Bay.
  \newblock {\em Phys. Rev. Lett.}, 121 (2018) 241805.
  \newblock \url{https://journals.aps.org/prl/abstract/10.1103/PhysRevLett.121.241805}.
  
\bibitem{Reno}
  G.~Bak et al., (RENO Collaboration).
  \newblock Measurement of Reactor Antineutrino Oscillation Amplitude and Frequency at RENO.
  \newblock {\em Phys. Rev. Lett.}, 121 (2018) 201801
  \newblock \url{https://journals.aps.org/prl/abstract/10.1103/PhysRevLett.121.201801}.

  \bibitem{Mueller}
  Th.~A.~Mueller, et~al.,
  \newblock Improved predictions of reactor antineutrino spectra.
  \newblock {\em Phys. Rev. C}, 83 (2011) 054615.
  \newblock \url{https://link.aps.org/doi/10.1103/PhysRevC.83.054615}.

\bibitem{Huber}
  P.~Huber,
  \newblock Determination of antineutrino spectra from nuclear reactors.
  \newblock {\em Phys. Rev. C}, 84 (2011) 024617.
  \newblock \url{http://link.aps.org/doi/10.1103/PhysRevC.84.024617}.

  \bibitem{Mention}
  G.~Mention, et~al.,
  \newblock Reactor antineutrino anomaly.
  \newblock {\em Phys. Rev. D}, 83 (2011) 073006.
  \newblock \url{http://link.aps.org/doi/10.1103/PhysRevD.83.073006}.

\bibitem{Vogel}
  P.~Vogel.,
  \newblock Analysis of the Anti-neutrino Capture on Protons.
  \newblock {\em Phys. Rev. D}, 29 (1984) 1918.
  \newblock \url{https://link.aps.org/doi/10.1103/PhysRevD.29.1918}.

\bibitem{DBay_bump}
  F.~P.~An et al., (Daya Bay Collaboration).
  \newblock Measurement of the Reactor Antineutrino Flux and Spectrum at Daya Bay.
  \newblock {\em Phys. Rev. Lett.}, 116 (2016) 061801.
  \newblock \url{https://journals.aps.org/prl/abstract/10.1103/PhysRevLett.116.061801}.

\bibitem{DChooz_bump}
  Y.~Abe, et~al., ( The Double Chooz Collaboration).
  \newblock Improved measurements of the neutrino mixing angle $\mathrm{\theta_{13}}$ with the Double Chooz detector.
  \newblock {\em Journal of High Energy Physics}, 74 (2015).
  \newblock \url{https://link.springer.com/article/10.1007/JHEP02(2015)074}.

  \bibitem{Sonzogni}
  A.~A.~Sonzogni, et~al.,
  \newblock Effects of Fission Yield Data in the Calculation of AntiNeutrino Spectra for $\mathrm{{}^{235}u(n,fission)}$ at Thermal and Fast Neutron Energies.
  \newblock {\em Phys. Rev. Lett.}, 116 (2016) 132502.
  \newblock \url{https://link.aps.org/doi/10.1103/PhysRevLett.116.132502}.
  
\bibitem{SONGS}
  N.~S.~Bowden, et~al.,
  \newblock Experimental results from an antineutrino detector for cooperative monitoring of nuclear reactors.
  \newblock {\em Nuclear Instruments and Methods in Physics Research Section A }, 572 (2007) 985.
  \newblock \url{http://www.sciencedirect.com/science/article/pii/S0168900206024326}.
  
\bibitem{NUCIFER}
  G.~Boireau, et~al., (Nucifer Collaboration).
  \newblock Online monitoring of the Osiris reactor with the Nucifer neutrino detector.
  \newblock {\em Phys. Rev. D}, 93 (2016) 112006.
  \newblock \url{https://link.aps.org/doi/10.1103/PhysRevD.93.112006}.

 \bibitem{DHRUVA}
  S.~K.~Agarwal, et~al.,
  \newblock Dhruva: Main design features, operational experience and utilization.
  \newblock{\em Nuclear Engineering and Design}, 236 (2006) 747-757.
  \newblock \url{https://www.sciencedirect.com/science/article/pii/S0029549306000732}.

 \bibitem{NEOS}
  Y.~J.~Ko, et~al., (NEOS Collaboration), 
  \newblock Sterile Neutrino Search at the NEOS Experiment. 
  \newblock{\em Phys. Rev. Lett.}, 118 (2017) 121802.
  \newblock \url{https://doi.org/10.1103/PhysRevLett.118.121802}.

 \bibitem{RENOST}
  J.~H.~Choi, et~al., (RENO Collaboration). 
  \newblock Search for Sub-eV Sterile Neutrinos at RENO.
  \newblock {\em Phys. Rev. Lett.}, 125 (2020) 191801.
  \newblock \url{https://journals.aps.org/prl/abstract/10.1103/PhysRevLett.125.191801}.
  
\bibitem{Shiba}
  S.~P.~Behera, D.~K.~Mishra and L.~M.~Pant,
  \newblock Active-sterile neutrino mixing constraints using reactor antineutrinos with the ISMRAN setup, 
  \newblock {Physical Review D}, 102 (2020) 013002.
  \newblock \url{https://journals.aps.org/prd/abstract/10.1103/PhysRevD.102.013002}. 

  
  \bibitem{PROSPECT}
  J~Ashenfelter, et~al., (PROSPECT Collaboration).
  \newblock Measurement of the Antineutrino Spectrum from ${}^{235}$U Fission at HFIR with PROSPECT.
  \newblock {\em Phys. Rev. Lett.}, 122 (2019) 251801.
  \newblock \url{https://doi.org/10.1103/PhysRevLett.122.251801}.

 \bibitem{DayaBayFuel}
  D.~Adey, et~al., (Daya Bay Collaboration),
  \newblock Extraction of the ${}^{235}$U and ${}^{239}$Pu Antineutrino Spectra at Daya Bay.
  \newblock {\em Phys. Rev. Lett.}, 123 (2019) 111801.
  \newblock \url{https://doi.org/10.1103/PhysRevLett.123.111801}.
  
\bibitem{IAEA}
  \newblock Technical Meeting on Nuclear Data for Anti-neutrino Spectra and Their Applications, 23-26 April 2019, IAEA Headquarters, Vienna, Austria.
  \newblock \url{https://www-nds.iaea.org/index-meeting-crp/Antineutrinos/}.
  
\bibitem{Oguri}
  S.~Oguri. et~al., (PANDA Collaboration),
  \newblock Reactor antineutrino monitoring with a plastic scintillator array as a new safeguards method.
  \newblock {\em Nuclear Instruments and Methods in Physics Research Section A:Accelerators, Spectrometers, Detectors and Associated Equipment}, 757 (2014) 33-39.
  \newblock \url{https://www.sciencedirect.com/science/article/abs/pii/S0168900214004781}.
  
  \bibitem{DB5MeV}
  F.~P.~An, et~al., (Daya Bay Collaboration), 
  \newblock Evolution of the Reactor Antineutrino Flux and Spectrum at Daya Bay.
  \newblock{\em Phys. Rev. Lett.},  118 (2017) 251801.
  \newblock \url{https://journals.aps.org/prl/abstract/10.1103/PhysRevLett.118.251801}. 

\bibitem{RENO5MeV}
  G.~Bak, et~al., (RENO Collaboration), 
  \newblock Fuel-Composition Dependent Reactor Antineutrino Yield at RENO.
  \newblock{\em Phys. Rev. Lett.}, 122 (2019) 232501.
  \newblock \url{https://journals.aps.org/prl/abstract/10.1103/PhysRevLett.122.232501}.

\bibitem{HUBER5MeV}
  P.~Huber, et~al., 
  \newblock NEOS Data and the Origin of the 5 MeV Bump in the Reactor Antineutrino Spectrum.
  \newblock {\em Phys. Rev. Lett.}, 118 (2017) 042502.
  \newblock \url{https://journals.aps.org/prl/abstract/10.1103/PhysRevLett.118.042502}.
  
\bibitem{miniISMRAN}
  P.~K~Netrakanti, et~al., (ISMRAN Collaboration), 
  \newblock Measurements using a prototype array of plastic scintillator bars for reactor base electron anti-neutrino detection.
  \newblock {\em Nuclear Instruments and Methods in Physics Research Section A: Accelerators, Spectrometers, Detectors and Associated Equipment}, 1024 (2022) 166126.
  \newblock \url{https://www.sciencedirect.com/science/article/pii/S016890022100992X}.
  
\bibitem{ISMRAN}
  D.~Mulmule, et~al., (ISMRAN Collaboration), 
  \newblock A plastic scintillator array for reactor based anti-neutrino studies.
  \newblock {\em Nuclear Instruments and Methods in Physics Research Section A: Accelerators, Spectrometers, Detectors and Associated Equipment}, 911 (2018) 104-114.
  \newblock \url{https://www.sciencedirect.com/science/article/abs/pii/S0168900218313408#!}.
  
  
\bibitem{DANSS}
  I.~Alekseev, et~al., (DANSS Collaboration).
  \newblock Danss: Detector of the reactor Antineutrino based on Solid Scintillator.
  \newblock {\em Journal of Instrumentation}, 11 (2016) P11011.
  \newblock \url{http://stacks.iop.org/1748-0221/11/i=11/a=P11011}.
  
\bibitem{eljen}
  {ELJEN}.
  \newblock \url{https://eljentechnology.com}
 
\bibitem{GEANT4}
  S.~Agostinelli, et~al.,
  \newblock GEANT4 - a simulation toolkit.
  \newblock {\em Nuclear Instruments and Methods in Physics Research Section A}, 506 (2003) 250-303.
  \newblock \url{http://www.sciencedirect.com/science/article/pii/S0168900203013688}.
  
\bibitem{DICEBOX}
  F.~Becvar, 
  \newblock Simulation of $\gamma$ cascades in complex nuclei with emphasis on assessment of uncertainties 
  of cascade-related quantities.
  \newblock {\em Nuclear Instruments and Methods in Physics Research Section A}, 417 (1998) 434-449. 
  \newblock \url{https://www.sciencedirect.com/science/article/abs/pii/S0168900298007876}.

 \bibitem{ANRIGD}
  T.~Tanaka, et~al,
  \newblock Gamma-ray spectra from thermal neutron capture on gadolinium-155 and natural gadolinium.
  \newblock {\em Prog.  Theor.  Exp.  Phys.}, 4 (2020) 043D02. 
  \newblock \url{https://doi.org/10.1093/ptep/ptaa015}.
  
\bibitem{RAMAN}
  R.~Sehgal, et.~al.
  \newblock A new technique to enhance the position resolution of large area plastic scinitillators to reconstruct the cosmic muon tracks.
  \newblock {\em Journal of Instrumation}, 17 (2022), P02036.
  \newblock \url{https://iopscience.iop.org/article/10.1088/1748-0221/17/02/P02036}.
  
\bibitem{MLP}
  D.~Mulmule, P.~K.~Netrakanti, L.~M.~Pant and B.~K.~Nayak. 
  \newblock Machine learning technique to improve anti-neutrino detection efficiency for the ISMRAN experiment.
  \newblock {\em Journal of Instrumation}, 15 (2020), P04021.
  \newblock \url{https://iopscience.iop.org/article/10.1088/1748-0221/15/04/P04021}.
  
\bibitem{Roni}
  R.~Dey, et.~al. (ISMRAN Collaboration)
  \newblock Characterization of plastic scintillator bars using fast neutrons from $D$-$D$ and $D$-$T$ reactions.
  \newblock {\em Journal of Instrumation}, 16 (2021), P08029.
  \newblock \url{https://iopscience.iop.org/article/10.1088/1748-0221/16/08/P08029}.
  
\bibitem{stereo}
  H.~Almazán, et.~al. (STEREO Collaboration)
  \newblock Improved sterile neutrino constraints from the STEREO experiment with 179 days of reactor-on data.
  \newblock {Phys. Rev. D}, 102, 052002.
  \newblock \url{https://journals.aps.org/prd/abstract/10.1103/PhysRevD.102.052002}.
  
  \bibitem{daya_bay}
  D.~Adey, et.~al. (Daya Bay Collaboration)
  \newblock A high precision calibration of the nonlinear energy response at Daya Bay.
  \newblock {\em Nuclear Instruments and Methods in Physics Research Section A}, 940 (2019) 230-242.
  \newblock \url{https://www.sciencedirect.com/science/article/abs/pii/S016890021930871X}.

  \bibitem{ReactorBkg}
  J.~Ashenfelter, et.~al., (The PROSPECT Collaboration),
  \newblock Background Radiation Measurements at High Power Research Reactors.
  \newblock {\em Nuclear Instruments and Methods in Physics Research Section A: Accelerators, Spectrometers, Detectors and Associated Equipment}, 806 (2016) 401-419.
  \newblock \url{https://doi.org/10.1016/j.nima.2015.10.023}.

  \bibitem{B12}
  C.~Galbiati and J~F.~Beacom
  \newblock Measuring the cosmic ray muon-induced fast neutron spectrum by (n,p) isotope production reactions in underground detectors.
  \newblock {\em Phys. Rev. C}, 72 (2005) 025807.
  \newblock \url{https://doi.org/10.1103/PhysRevC.72.025807}.

  \bibitem{cosmic}
  S.P.Behera, et.~al. 
  \newblock Cosmic ray measurements using the ISMRAN setup in a non-reactor environment.
  \newblock {\em Astroparticle Physics}, 141 (2022) 102729.
  \newblock \url{https://doi.org/10.1016/j.astropartphys.2022.102729}.
  
  \bibitem{danss_expt}
  D.Svirida, et.~al. (The DANSS Collaboration)
  \newblock DANSS experiment: current status and future plans.
  \newblock {J. Phys.: Conf. Ser.} (2020) 1690 012179.
  \newblock \url{https://iopscience.iop.org/article/10.1088/1742-6596/1690/1/012179}.
  
 \bibitem{cosmic_flux}
  {Particle Data Group}.
  \newblock \url{https://pdg.lbl.gov/2011/reviews/rpp2011-rev-cosmic-rays.pdf}

  \bibitem{reno_tof1}
  C.~D.~Shin, et~al. (The RENO Collaboration)
  \newblock Observation of reactor antineutrino disappearance using delayed neutron capture on hydrogen at RENO.
  \newblock {\em The Journal of High Energy Physics}, 29 (2020).
  \newblock \url{https://doi.org/10.1007/JHEP04(2020)029}.

  \bibitem{reno_tof2}
  S.~H.~Seo, et~al. (RENO Collaboration)
  \newblock Spectral measurement of the electron antineutrino oscillation amplitude and frequency using 500 live days of RENO data.
  \newblock {\em Physical Review D}, 98 (2018) 012002.
  \newblock \url{https://journals.aps.org/prd/abstract/10.1103/PhysRevD.98.012002}.

  \bibitem{Lorch}
  Edgar.~A.~Lorch, et~al.,
  \newblock Neutron Spectra of ${}^{241}Am$/B, ${}^{241}Am$/Be, ${}^{241}Am$/F, ${}^{242}Cm$/Be, ${}^{238}Pu/{}^{13}C$ and ${}^{252}Cf$ isotopic neutron sources.
  \newblock {\em The International journal of Applied Radiation and Isotopes}, 24 (1973).
  \newblock \url{https://doi.org/10.1016/0020-708X(73)90127-0}.

  \bibitem{tof_para}
  R.~A.~Cecil, et~al.
  \newblock Improved predections of neutron detection efficiency for hydrocarbon scintillators from 1 MeV to about 300 MeV.
  \newblock {\em Nuclear Instruments and Methods in Physics Research Section A: Accelerators, Spectrometers, Detectors and Associated Equipment}, 161 (1979) 439-447.
  \newblock \url{https://www.sciencedirect.com/science/article/abs/pii/0029554X79904178}.



  

\end{thebibliography}
\end{document}